\documentstyle{article}
\begin{document}

\title{ An efficient approach for spin-angular integrations in atomic structure
calculations}
\author{Gediminas Gaigalas, Zenonas Rudzikas \\
%EndAName
{\em State Institute of Theoretical Physics and Astronomy,} \\
{\em A. Go\v stauto 12, 2600 Vilnius, LITHUANIA} \\
\ \\
Charlotte Froese Fischer \\
{\em Department of Computer Science, Box 1679B,} \\
{\em Vanderbilt University, Nashville, TN 37235, USA}}
\date{}
\maketitle

\vspace{.5in} An approach for spin-angular integrations

\vspace{.5in} {\bf PACS: 0270, 3110, 3115}

\clearpage

\begin{abstract}

  A general method is described for finding algebraic expressions for
 matrix elements of any one- and two-particle operator for an arbitrary
 number of subshells in an atomic configuration, requiring
   neither coefficients of fractional parentage
   nor unit tensors.
 It is based on the combination of
   second quantization in the coupled tensorial form,
   angular momentum theory in three spaces (orbital, spin and quasispin),
   and a generalized graphical technique.
  The latter allows us to
   calculate graphically the irreducible tensorial products of the
    second quantization operators and their commutators,
   and to formulate additional rules for operations with diagrams.
 The additional rules allow us
   to find graphically the normal form of the complicated tensorial products
    of the operators.
 All matrix elements (diagonal and non-diagonal with respect to
 configurations)
   differ only by the values of the projections of the
     quasispin momenta of separate shells
   and are expressed in terms of completely reduced  matrix
 elements (in all three spaces) of the second quantization operators.
As a result,  it allows us to use standard quantities uniformly for both diagonal
and  off-diagonal matrix elements.
%GG
%Preliminary calculations indicate that computer programs based on
%this approach on average are 4-6 times faster than other well-known
%angular codes.
\end{abstract}

\clearpage

\section{\bf Introduction}

In order to obtain accurate values of atomic quantities it is necessary to
account for relativistic and correlation effects. Relativistic effects may
be taken into account as Breit-Pauli corrections or, in a fully
relativistic approach, by
starting with the Dirac-Coulomb Hamiltonian and wave functions
defined in terms of four-component one-electron orbitals.
In both cases, correlation effects may be considered either variationally
or perturbatively. For complex atoms and ions, a
considerable part of the effort must be devoted to coping with integrations
over spin-angular variables, occurring in the matrix elements of the
operators under consideration.

Many existing codes for integrating the spin-angular parts of matrix elements
(Glass 1978, Glass and Hibbert 1978, Grant 1988, Burke {\it et al }1994) are
based on the computational scheme proposed by Fano (1965). In  essence,
it consists of evaluating recoupling matrices. Although such an
approach uses Racah algebra, it may be necessary to
carry out multiple summations over intermediate terms. Due to
these summations and the complexity of the recoupling matrix
itself, the associated computer codes become rather time consuming. A
solution to this problem was found by Burke {\it et al } (1994). They tabulated
separate standard parts of recoupling matrices along with coefficients of
fractional parentage at the beginning of a calculation and used them further
on to calculate the needed coefficients. Computer codes by Glass (1978),
Glass and Hibbert (1978), Grant (1988), Burke {\it et al }(1994) utilize the
program NJSYM (Burke 1970) or NJGRAF (Bar-Shalom and Klapisch 1988) for
the calculation of recoupling matrices. Both are rather time consuming
when calculating matrix elements of complex operators or
electronic configurations with many open subshells.

In order to simplify the calculations, Cowan (1981) suggested that
matrix elements be grouped into ''Classes'' (see Cowan 1981 Figure 13-5).
Unfortunately, this approach was not generalized to all two-electron operators.
Perhaps for this reason Cowan's approach is not widely used
although the program itself, based on this approach is widely used.

Many approaches for the calculation of spin-angular coefficients (Glass
1978, Glass and Hibbert 1978, Grant 1988, Burke {\it et al }1994) are based
on the usage of Racah algebra only on the level of coefficients of
fractional parentage. A few authors (Jucys and Savukynas 1973, Cowan 1981)
utilize the unit tensors, simplifying the calculations in this way, because
use can be made of the tables of unit tensors and selection rules
can be used to check whether the spin-angular coefficients
are zero prior to computation. Moreover, the recoupling
matrices themselves have a simpler form. Unfortunately, these ideas were
applied only to diagonal matrix elements with respect to configurations,
though Cowan (1981) suggested the usage of unit tensors for
non-diagonal ones as well.

All the above mentioned approaches were applied in the coordinate
representation. The second quantization formalism (Judd
1967, Rudzikas and Kaniauskas 1984, Rudzikas 1991 and Rudzikas 1997) has a
number of advantages compared to coordinate representation. First of all, it
is much easier to find algebraic expressions for complex operators and
their matrix elements, when relying on second quantization formalism. It has
contributed significantly to the successful development of
perturbation theory (see Lindgren and Morrison 1982, Merkelis {\it et al}
1985), and orthogonal operators (Uylings 1984), where three-particle
operators already occur. Uylings (1992) suggested a fairly simple approach
for dealing with separate cases of three-particle operators. Moreover, in the
second quantization approach the quasispin formalism was efficiently
developed by Rudzikas and Kaniauskas (1984). The main advantage
of this approach is
that applying the quasispin
method for calculating the matrix elements of any operator, we
can use the reduced coefficients of fractional parentage
whose  matrix elements are independent of the
occupation number of the shell.
All this enabled Merkelis and
Gaigalas (1985) to work out a general perturbation theory approach for complex
cases of several open shells.

Thus, it seems that it is possible to
formulate an efficient and general
approach for finding the spin-angular parts of matrix elements of
atomic interactions, relying on the combination of the second
quantization approach in the
coupled tensorial form, the generalized graphical technique and
angular momentum theory in orbital, spin and quasispin spaces
as well as the symmetry properties of the quantities
considered, which would
be free of previous shortcomings. Gaigalas and Rudzikas (1996)
suggested such an approach  for finding matrix elements of any
one- and two-particle atomic operator for the case of two open
shells of equivalent electrons.
But the situation is different when the matrix elements
between more complex configurations are considered.

An approach for the latter case  is described in the present paper.
One of the main ideas proposed here allows one to solve the problems
related to the more complex configurations. Namely, we propose to apply
Wick's theorem (see Lindgren and Morrison 1982) not in
its usual general form while calculating the matrix elements, but
rather only for groups of operators acting upon distinct shells of
equivalent electrons. So, the ordering of operators generally would
not be normal.

It is a universal approach for finding
algebraic expressions for matrix elements of any one- and two-particle
operator in the general case of an arbitrary number of subshells in
an atomic configuration, heavily based on the exploitation of the quasispin
technique and of the Wigner-Eckart theorem in quasispin space.
Expressions for matrix elements and
recoupling matrices are obtained by first classifying each matrix element
into one of four classes, depending on the number of subshells
being acted upon. Each class is then further subdivided into
cases and explicit expressions derived for each case in terms
of triangular conditions, $nj$-symbols, and
reduced matrix elements.  From these expressions, efficient
procedures can be developed, that apply tests before performing
computation.

Tensorial expressions for any two-particle operator are presented in
Section 2. They are based on the underlying assumption that the second
quantization operators (both creation and annihilation),
acting on the same open shell, must always be beside one another in
a
tensorial product and must be coupled into a resultant momentum. Then the
second quantization operators, acting on the next shell, must follow, etc.
Section 3 deals with the matrix elements between complex configurations.
General expressions for recoupling matrices were found (Section 4)
by use of the modified graphical technique of Jucys and Bandzaitis (1977),
allowing us to calculate graphically the irreducible tensorial
products of the second quantization operators and their
commutators, and to formulate additional rules for operations
with diagrams. The additional rules allow us to find
graphically the normal form of the complicated tensorial
products of the operators. All the graphical transformations
we use here are fully described in Gaigalas and Rudzikas 1996.

Exploitation of this new version of Racah algebra based on the
angular momentum theory, on a generalized graphical approach,
on quasispin approach, and on the use of reduced coefficients
of fractional parentage
for finding the spin-angular parts of two-particle operators is
outlined in Sections 5-7. Some details of the calculations are presented in
the appendix.

\section{\bf Tensorial expressions for any two-particle operators}

In order to be able to find the expressions for matrix elements
of the operators studied, we have to express these operators in
terms of the irreducible tensors or their irreducible products.
In this section we will present all the necessary tensorial expressions for any
two-particle operator $G$.

First, we express the operator in second-quantization
form (Gaigalas and Rudzikas 1996) as

\begin{equation}
\label{eq:pirga}G=\displaystyle {\sum_{n_il_i,n_jl_j,n_i^{\prime
}l_i^{\prime },n_j^{\prime }l_j^{\prime }}}\widehat{G}(n_il_in_jl_jn_i^{\prime
}l_i^{\prime }n_j^{\prime }l_j^{\prime })=\\\frac 12\displaystyle
{\sum_{i,j,i^{\prime },j^{\prime }}}a_ia_ja_{j^{\prime }}^{\dagger}
a_{i^{\prime
}}^{\dagger}\left( i,j|g|i^{\prime },j^{\prime }\right)
\end{equation}
where, as is customary, the creation operators $a_ia_j$ appear
to the left of the annihilation operators
$a_{j^{\prime}}^{\dagger}a_{i^{\prime}}^{\dagger}$
before defining the shells upon which the second quantization operators
are acting. After defining the shells explicitly, the second quantization
operators are transformed using their commutation relations so that all
operators with the same $n\lambda $ ($\lambda \equiv l,s$) are beside
one another. %CFF  Is order important -- before? after? both can be beside.
For example, in the case where the electron creation operator $a_i,$ and electron
annihilation operators $a_{i^{\prime }}^{\dagger}$ (where $i\equiv
n_il_ism_{l_i}m_{s_i}$) and $a_{j^{\prime }}^{\dagger}$ act upon the same shell $%
\alpha $, and operator $a_j$ acts upon another shell $\beta $, we have:
%CFF What is G  -- any general operator?
%CFF Does G not depend on \kappa_1, kappa2, k, \sigma_1, \sigma_2 ??
%CFF How does (kk) couple?  Is it arbitrary?
\begin{equation}
\label{eq:ma}
\begin{array}{c}
\widehat{G}\left( \alpha \beta \alpha \alpha \right) =\frac 12
\displaystyle {\sum_{m_{l_\alpha }m_{s_\alpha }m_{l_\beta }m_{s_\beta }}}%
\displaystyle {\sum_{m_{l_\alpha }^{\prime }m_{s_\alpha }^{\prime
}m_{l_\alpha }^{\prime \prime }m_{s_\alpha }^{\prime \prime }}}%
a_{m_{l_\alpha }m_{s_\alpha }}^{\left( l_\alpha s_\alpha \right)
}a_{m_{l_\alpha }^{\prime \prime }m_{s_\alpha }^{\prime \prime }}^{\dagger
\left(
l_\alpha s_\alpha \right) }a_{m_{l_\alpha }^{\prime }m_{s_\alpha }^{\prime
}}^{\dagger \left( l_\alpha s_\alpha \right) }a_{m_{l_\beta }m_{s_\beta
}}^{(l_\beta s_\beta )}\times \\ \times \left( n_\alpha \lambda _\alpha
m_{l_\alpha }m_{s_\alpha }n_\beta \lambda _\beta m_{l_\beta }m_{s_\beta
}|g^{\left( \kappa _1\kappa _2k,\sigma _1\sigma _2k\right) }|n_\alpha
\lambda _\alpha m_{l_\alpha }^{\prime }m_{s_\alpha }^{\prime }n_\alpha
\lambda _\alpha m_{l_\alpha }^{\prime \prime }m_{s_\alpha }^{\prime \prime
}\right) .
\end{array}
\end{equation}
Here we imply that a tensorial structure indexed by $%
(\kappa_1\kappa_2k,\sigma _1\sigma _2 k )$ at $g$ has rank $\kappa_1$
for electron $1$, rank $\kappa_2$ for electron $2$, and a resulting
rank $k$ in the $l$ space, and
corresponding ranks $\sigma _1\sigma _2 k $ in the $s$ space.

Now, applying the graphical approach of angular momentum theory (Gaigalas
{\it et al} 1985), we can get an expression, namely

\begin{equation}
\label{eq:mb}
\begin{array}[b]{c}
\widehat{G}\left( \alpha \beta \alpha \alpha \right) =\frac 12
\displaystyle {\sum_{\kappa _{12}\sigma _{12}\kappa _{12}^{\prime }\sigma
_{12}^{\prime }}}\displaystyle {\sum_p}\left( -1\right) ^{\kappa
_{12}+\sigma _{12}-\kappa _{12}^{\prime }-\sigma _{12}^{\prime }-k-p}\times
\\ \times \left( n_\alpha \lambda _\alpha n_\beta \lambda _\beta ||g^{\left(
\kappa _1\kappa _2k,\sigma _1\sigma _2k\right) }||n_\alpha \lambda _\alpha
n_\alpha \lambda _\alpha \right) \times \\
\times \left[ \kappa _{12},\sigma _{12}\right] \left[ \kappa _{12}^{\prime
},\sigma _{12}^{\prime }\right] ^{1/2}\left\{
\begin{array}{ccc}
l_\alpha & l_\alpha & \kappa _{12}^{\prime } \\
\kappa _1 & \kappa _2 & k \\
l_\alpha & l_\beta & \kappa _{12}
\end{array}
\right\} \left\{
\begin{array}{ccc}
s & s & \sigma _{12}^{\prime } \\
\sigma _1 & \sigma _2 & k \\
s & s & \sigma _{12}
\end{array}
\right\} \times \\
\times \displaystyle {\sum_{K_lK_s}}\left[ K_l,K_s\right] ^{1/2}\left\{
\begin{array}{ccc}
l_\beta & l_\alpha & \kappa _{12} \\
\kappa _{12}^{\prime } & k & K_l
\end{array}
\right\} \left\{
\begin{array}{ccc}
s & s & \sigma _{12} \\
\sigma _{12}^{\prime } & k & K_s
\end{array}
\right\} \times \\
\times \left[ a^{\left( l_\beta s\right) }\times \left[ a^{\left( l_\alpha
s\right) }\times \left[ \stackrel{\sim }{a}^{\left( l_\alpha s\right)
}\times \stackrel{\sim }{a}^{\left( l_\alpha s\right) }\right] ^{\left(
\kappa _{12}^{\prime }\sigma _{12}^{\prime }\right) }\right] ^{\left(
K_lK_s\right) }\right] _{p,-p}^{\left( kk\right) }
\end{array}
\end{equation}
where $\left[ a,b\right] =\left( 2a+1\right) \left( 2b+1\right) $,
$\left( n_\alpha \lambda _\alpha n_\beta \lambda _\beta ||g^{\left(
\kappa _1\kappa _2k,\sigma _1\sigma _2k\right) }||
n_\alpha \lambda _\alpha n_\alpha \lambda _\alpha \right)$ is
the two-electron submatrix (reduced matrix) element of operator
$\widehat{G}$
and $\stackrel{\sim }{a}^{\left(l_{\alpha} s\right) }$ is defined as
(Judd 1967)

\begin{equation}
\label{eq:ffggc}
\stackrel{\sim}{a}_{m_{l_\alpha} m_s}^{\left(l_{\alpha} s \right)
}=\left( -1\right)^{l_\alpha+s-m_{l_\alpha}-m_s }a_{-m_{l_\alpha}-m_s }
^{\dagger \left(l_{\alpha} s \right)}.
\end{equation}
Expression (\ref{eq:mb}) has summations over intermediate
ranks $ \kappa _{12}^{\prime }, \sigma
_{12}^{\prime }, K_l, K_s$ in tensorial product. The angular
momentum projection of $(kk)$ is $p,-p$.

In order to calculate the spin-angular part of a two-particle operator
matrix element with an arbitrary number of open shells, it is necessary to
consider all possible distributions of subshells, upon which the
second quantization operators are acting. These are presented in
Table \ref{pasis}.
We point  out that for distributions 2 -- 5 and 19 -- 42 the shells'
sequence numbers $\alpha $, $\beta $, $\gamma $, $\delta $ (in bra and ket
functions of a submatrix element) satisfy the condition $\alpha <\beta <\gamma
<\delta $, while for distributions 6 -- 18 no conditions upon $\alpha $%
, $\beta $, $\gamma $, $\delta $ are imposed.

\clearpage
\begin{table}
\begin{center}
\caption{Distributions of subshells, upon which the second quantization
operators are acting, that appear in the submatrix elements of any
two-particle operator, when bra and ket functions have $u$ open
subshells}
\label{pasis}
  \begin{tabular}{|r|c|c|c|c|c|} \hline
     &  &  &  &  & \\
    {\bf No.} & {\bf $a_{i}$ } & {\bf $a_{j}$ } & {\bf $a_{i'}^{\dagger}$ } &
    {\bf $a_{j'}^{\dagger}$ }
& {\bf submatrix element} \\
     &  &  &  &  & \\ \hline \hline
{\bf  1.} & $\alpha$  & $\alpha$ &  $\alpha$ & $\alpha$
& $ (...n_{\alpha}l_{\alpha}^{N_{\alpha}}...||
\widehat{G}(n_il_in_jl_jn_i^{\prime}l_i^{\prime }n_j^{\prime}l_j^{\prime })
||...n_{\alpha}l_{\alpha}^{N_{\alpha}}... )$ \\  \hline  \hline
{\bf  2.} & $\alpha$  & $\beta$  &  $\alpha$ &  $\beta$  & \\
{\bf  3.} & $\beta$   & $\alpha$ & $\beta$  & $\alpha$
& $ (...n_{\alpha}l_{\alpha}^{N_{\alpha}}...n_{\beta}l_{\beta}^{N_{\beta}}...||$ \\
{\bf  4.} & $\alpha$  & $\beta$  & $\beta$  & $\alpha$ &
$\widehat{G}
(n_il_in_jl_jn_i^{\prime}l_i^{\prime }n_j^{\prime}l_j^{\prime })$ \\
{\bf  5.} & $\beta$   & $\alpha$ & $\alpha$ & $\beta$
&$||...n_{\alpha}l_{\alpha}^{N_{\alpha}}...n_{\beta}l_{\beta}^{N_{\beta}}...)$ \\    \hline
{\bf  6.} & $\alpha$ & $\alpha$  & $\beta$  & $\beta$
& $ (...n_{\alpha}l_{\alpha}^{N_{\alpha}}...n_{\beta}l_{\beta}^{N_{\beta}}...||
\widehat{G}
||...n_{\alpha}l_{\alpha}^{N_{\alpha}-2}...n_{\beta}l_{\beta}^{N_{\beta}+2}...)$ \\    \hline
{\bf  7.} & $\beta$    & $\alpha$ & $\alpha$ & $\alpha$  & \\
{\bf  8.} & $\alpha$   & $\beta$  & $\alpha$ & $\alpha$
& $(...n_{\alpha}l_{\alpha}^{N_{\alpha}}...n_{\beta}l_{\beta}^{N_{\beta}}...||$ \\
{\bf  9.} & $\beta$    & $\beta$  & $\beta$  & $\alpha$  &
$\widehat{G}
(n_il_in_jl_jn_i^{\prime}l_i^{\prime }n_j^{\prime}l_j^{\prime })$ \\
{\bf 10.} & $\beta$   & $\beta$  & $\alpha$ & $\beta$
& $||...n_{\alpha}l_{\alpha}^{N_{\alpha}+1}...n_{\beta}l_{\beta}^{N_{\beta}-1}...)$ \\    \hline
 \hline
{\bf 11.} & $\beta$   & $\gamma$ & $\alpha$ & $\gamma$  & \\
{\bf 12.} & $\gamma$  & $\beta$  & $\gamma$ & $\alpha$
& $ (...n_{\alpha}l_{\alpha}^{N_{\alpha}}n_{\beta}l_{\beta}^{N_{\beta}}
n_{\gamma}l_{\gamma}^{N_{\gamma}}...||$ \\
{\bf 13.} & $\gamma$  & $\beta$  & $\alpha$ & $\gamma$  &
$\widehat{G}
(n_il_in_jl_jn_i^{\prime}l_i^{\prime }n_j^{\prime}l_j^{\prime })$  \\
{\bf 14.} & $\beta$   & $\gamma$ & $\gamma$ & $\alpha$
& $||...n_{\alpha}l_{\alpha}^{N_{\alpha}+1}n_{\beta}l_{\beta}^{N_{\beta}-1}
n_{\gamma}l_{\gamma}^{N_{\gamma}}...)$ \\    \hline
{\bf 15.} & $\gamma$  & $\gamma$ & $\alpha$ & $\beta$
& $ (...n_{\alpha}l_{\alpha}^{N_{\alpha}}n_{\beta}l_{\beta}^{N_{\beta}}
n_{\gamma}l_{\gamma}^{N_{\gamma}}...||
\widehat{G}
(n_il_in_jl_jn_i^{\prime}l_i^{\prime }n_j^{\prime}l_j^{\prime })$ \\
{\bf 16.} & $\gamma$  & $\gamma$ & $\beta$  & $\alpha$
& $||...n_{\alpha}l_{\alpha}^{N_{\alpha}+1}n_{\beta}l_{\beta}^{N_{\beta}+1}
n_{\gamma}l_{\gamma}^{N_{\gamma}-2}...)$ \\    \hline
{\bf 17.} & $\alpha$  & $\beta$  & $\gamma$ & $\gamma$
& $ (...n_{\alpha}l_{\alpha}^{N_{\alpha}}n_{\beta}l_{\beta}^{N_{\beta}}
n_{\gamma}l_{\gamma}^{N_{\gamma}}...||
\widehat{G}
(n_il_in_jl_jn_i^{\prime}l_i^{\prime }n_j^{\prime}l_j^{\prime })$ \\
{\bf 18.} & $\beta$   & $\alpha$ & $\gamma$ & $\gamma$
& $||...n_{\alpha}l_{\alpha}^{N_{\alpha}-1}n_{\beta}l_{\beta}^{N_{\beta}-1}
n_{\gamma}l_{\gamma}^{N_{\gamma}+2}...)$ \\    \hline
 \hline
{\bf 19.} & $\alpha$  & $\beta$  & $\gamma$ & $\delta$  & \\
{\bf 20.} & $\beta$   & $\alpha$ & $\gamma$ & $\delta$
& $ (n_{\alpha}l_{\alpha}^{N_{\alpha}}n_{\beta}l_{\beta}^{N_{\beta}}
n_{\gamma}l_{\gamma}^{N_{\gamma}}n_{\delta}l_{\delta}^{N_{\delta}}||$ \\
{\bf 21.} & $\alpha$  & $\beta$  & $\delta$ & $\gamma$  &
$\widehat{G}
(n_il_in_jl_jn_i^{\prime}l_i^{\prime }n_j^{\prime}l_j^{\prime })$ \\
{\bf 22.} & $\beta$   & $\alpha$ & $\delta$ & $\gamma$
& $||n_{\alpha}l_{\alpha}^{N_{\alpha}-1}n_{\beta}l_{\beta}^{N_{\beta}-1}
n_{\gamma}l_{\gamma}^{N_{\gamma}+1}n_{\delta}l_{\delta}^{N_{\delta}+1})$ \\    \hline
 \end{tabular}
\end{center}
\end{table}

\clearpage
\vspace {0.5in} Table 1 (continued) \vspace {0.1in}

\begin{center}
%\caption{Table 1 (continued)}
\begin{tabular}{|l|l|l|l|l|c|} \hline
     &  &  &  &  & \\
    {\bf No.} & {\bf $a_{i}$ } & {\bf $a_{j}$ } & {\bf $a_{i'}^{\dagger}$ }
     & {\bf $a_{j'}^{\dagger}$ }
& {\bf submatrix element} \\
     &  &  &  &  & \\ \hline \hline
{\bf 23.} & $\gamma$  & $\delta$ & $\alpha$ & $\beta$  & \\
{\bf 24.} & $\gamma$  & $\delta$ & $\beta$  & $\alpha$
& $(n_{\alpha}l_{\alpha}^{N_{\alpha}}n_{\beta}l_{\beta}^{N_{\beta}}
n_{\gamma}l_{\gamma}^{N_{\gamma}}n_{\delta}l_{\delta}^{N_{\delta}}||$ \\
{\bf 25.} & $\delta$  & $\gamma$ & $\alpha$ & $\beta$  &
$\widehat{G}
(n_il_in_jl_jn_i^{\prime}l_i^{\prime }n_j^{\prime}l_j^{\prime })$ \\
{\bf 26.} & $\delta$  & $\gamma$ & $\beta$  & $\alpha$
& $||n_{\alpha}l_{\alpha}^{N_{\alpha}+1}n_{\beta}l_{\beta}^{N_{\beta}+1}
n_{\gamma}l_{\gamma}^{N_{\gamma}-1}n_{\delta}l_{\delta}^{N_{\delta}-1})$ \\    \hline
{\bf 27.} & $\alpha$  & $\gamma$ & $\beta$  & $\delta$  & \\
{\bf 28.} & $\alpha$  & $\gamma$ & $\delta$ & $\beta$
& $ (n_{\alpha}l_{\alpha}^{N_{\alpha}}n_{\beta}l_{\beta}^{N_{\beta}}
n_{\gamma}l_{\gamma}^{N_{\gamma}}n_{\delta}l_{\delta}^{N_{\delta}}||$ \\
{\bf 29.} & $\gamma$  & $\alpha$ & $\delta$ & $\beta$  &
$\widehat{G}
(n_il_in_jl_jn_i^{\prime}l_i^{\prime }n_j^{\prime}l_j^{\prime })$ \\
{\bf 30.} & $\gamma$  & $\alpha$ & $\beta$  & $\delta$
& $||n_{\alpha}l_{\alpha}^{N_{\alpha}-1}n_{\beta}l_{\beta}^{N_{\beta}+1}
n_{\gamma}l_{\gamma}^{N_{\gamma}-1}n_{\delta}l_{\delta}^{N_{\delta}+1})$ \\    \hline
{\bf 31.} & $\beta$   & $\delta$ & $\alpha$ & $\gamma$  & \\
{\bf 32.} & $\delta$  & $\beta$  & $\gamma$ & $\alpha$
& $ (n_{\alpha}l_{\alpha}^{N_{\alpha}}n_{\beta}l_{\beta}^{N_{\beta}}
n_{\gamma}l_{\gamma}^{N_{\gamma}}n_{\delta}l_{\delta}^{N_{\delta}}||$ \\
{\bf 33.} & $\beta$   & $\delta$ & $\gamma$ & $\alpha$  &
$\widehat{G}
(n_il_in_jl_jn_i^{\prime}l_i^{\prime }n_j^{\prime}l_j^{\prime })$ \\
{\bf 34.} & $\delta$  & $\beta$  & $\alpha$ & $\gamma$
& $||n_{\alpha}l_{\alpha}^{N_{\alpha}+1}n_{\beta}l_{\beta}^{N_{\beta}-1}
n_{\gamma}l_{\gamma}^{N_{\gamma}+1}n_{\delta}l_{\delta}^{N_{\delta}-1})$ \\    \hline
{\bf 35.} & $\alpha$  & $\delta$ & $\beta$  & $\gamma$  & \\
{\bf 36.} & $\delta$  & $\alpha$ & $\gamma$ & $\beta$
& $ (n_{\alpha}l_{\alpha}^{N_{\alpha}}n_{\beta}l_{\beta}^{N_{\beta}}
n_{\gamma}l_{\gamma}^{N_{\gamma}}n_{\delta}l_{\delta}^{N_{\delta}}||$ \\
{\bf 37.} & $\alpha$  & $\delta$ & $\gamma$ & $\beta$  &
$\widehat{G}
(n_il_in_jl_jn_i^{\prime}l_i^{\prime }n_j^{\prime}l_j^{\prime })$ \\
{\bf 38.} & $\delta$  & $\alpha$ & $\beta$  & $\gamma$
& $||n_{\alpha}l_{\alpha}^{N_{\alpha}-1}n_{\beta}l_{\beta}^{N_{\beta}+1}
n_{\gamma}l_{\gamma}^{N_{\gamma}+1}n_{\delta}l_{\delta}^{N_{\delta}-1})$ \\    \hline
{\bf 39.} & $\beta$  & $\gamma$ & $\alpha$ & $\delta$  & \\
{\bf 40.} & $\gamma$ & $\beta$  & $\delta$ & $\alpha$
& $ (n_{\alpha}l_{\alpha}^{N_{\alpha}}n_{\beta}l_{\beta}^{N_{\beta}}
n_{\gamma}l_{\gamma}^{N_{\gamma}}n_{\delta}l_{\delta}^{N_{\delta}}||$ \\
{\bf 41.} & $\beta$  & $\gamma$ & $\delta$ & $\alpha$  &
$\widehat{G}
(n_il_in_jl_jn_i^{\prime}l_i^{\prime }n_j^{\prime}l_j^{\prime })$ \\
{\bf 42.} & $\gamma$ & $\beta$  & $\alpha$ & $\delta$
& $||n_{\alpha}l_{\alpha}^{N_{\alpha}+1}n_{\beta}l_{\beta}^{N_{\beta}-1}
n_{\gamma}l_{\gamma}^{N_{\gamma}-1}n_{\delta}l_{\delta}^{N_{\delta}+1})$ \\    \hline
  \end{tabular}
\end{center}

\vspace {0.5in}

Let $\Xi$ be an array of
intermediate coupling parameters in tensorial form, including
$\kappa _{12},\sigma _{12},\kappa _{12}^{\prime },\sigma
_{12}^{\prime}$ and possibly also others.  Then the tensorial
expressions for all these distributions can be grouped into four
classes, where in each class, the two-particle operator
$\widehat{G}$, operating on specific shells (see Eq. (1)),
has one of four forms:

\begin{enumerate}
\item  All the second quantization operators act upon the same shell
(distribution 1) and

\begin{equation}
\label{eq:mc}\widehat{G}\left( I\right) \sim \displaystyle
{\sum_{\kappa _{12},\sigma _{12},\kappa _{12}^{\prime },\sigma _{12}^{\prime
}}}\displaystyle {\sum_p}\Theta \left( n\lambda ,\Xi \right)
A_{p,-p}^{\left( kk\right) }\left( n\lambda ,\Xi \right) ;
\end{equation}

\item  The second quantization operators act upon the two different
shells (distributions 2-10) and

\begin{equation}
\label{eq:md}
\begin{array}[b]{c}
\widehat{G}\left( II\right) \sim
\displaystyle {\sum_{\kappa _{12},\sigma _{12},\kappa _{12}^{\prime },\sigma
_{12}^{\prime }}}\displaystyle {\sum_p}\Theta \left( n_\alpha \lambda
_\alpha ,n_\beta \lambda _\beta ,\Xi \right) \times  \\ \times \left[
B^{\left( \kappa _{12}\sigma _{12}\right) }\left( n_\alpha \lambda _\alpha
,\Xi \right) \times C^{\left( \kappa _{12}^{\prime }\sigma _{12}^{\prime
}\right) }\left( n_\beta \lambda _\beta ,\Xi \right) \right] _{p,-p}^{\left(
kk\right) };
\end{array}
\end{equation}

\item  The second quantization operators act upon three shells
(distributions 11-18)

\begin{equation}
\label{eq:me}
\begin{array}[b]{c}
\widehat{G}\left( III\right) \sim
\displaystyle {\sum_{\kappa _{12},\sigma _{12},\kappa _{12}^{\prime },\sigma
_{12}^{\prime }}}\displaystyle {\sum_p}\Theta \left( n_\alpha \lambda
_\alpha ,n_\beta \lambda _\beta ,n_\gamma \lambda _\gamma ,\Xi \right)
\times  \\ \times \left[ \left[ D^{\left( l_\alpha s\right) }\times
D^{\left( l_\beta s\right) }\right] ^{\left( \kappa _{12}\sigma _{12}\right)
}\times E^{\left( \kappa _{12}^{\prime }\sigma _{12}^{\prime }\right)
}\left( n_\gamma \lambda _\gamma ,\Xi \right) \right] _{p,-p}^{\left(
kk\right) };
\end{array}
\end{equation}

\item  The second quantization operators act upon four shells
(distributions 19-42) and

\begin{equation}
\label{eq:mf}
\begin{array}[b]{c}
\widehat{G}\left( IV\right) \sim
\displaystyle {\sum_{\kappa _{12},\sigma _{12},\kappa _{12}^{\prime },\sigma
_{12}^{\prime }}}\displaystyle {\sum_p}\Theta \left( n_\alpha \lambda
_\alpha ,n_\beta \lambda _\beta ,n_\gamma \lambda _\gamma ,n_\delta \lambda
_\delta ,\Xi \right) \times  \\ \times \left[ \left[ D^{\left( l_\alpha
s\right) }\times D^{\left( l_\beta s\right) }\right] ^{\left( \kappa
_{12}\sigma _{12}\right) }\times \left[ D^{\left( l_\gamma s\right) }\times
D^{\left( l_\delta s\right) }\right] ^{\left( \kappa _{12}^{\prime }\sigma
_{12}^{\prime }\right) }\right] _{p,-p}^{\left( kk\right) }.
\end{array}
\end{equation}
\end{enumerate}

In (\ref{eq:mc})-(\ref{eq:mf}), $\Theta \left( n\lambda ,\Xi \right) $,...,$%
\Theta \left( n_\alpha \lambda _\alpha ,n_\beta \lambda _\beta ,n_\gamma
\lambda _\gamma ,n_\delta \lambda _\delta ,\Xi \right) $ are proportional to
the radial part of the operator $\widehat{G}$, and
$A^{\left( kk\right) }\left(
n\lambda ,\Xi \right) $,...,$E^{\left( kk'\right) }\left( n\lambda ,\Xi
\right) $ denote tensorial products of irreducible tensors. Parameter
$\Xi$
implies the array of coupling parameters that connect $\Theta $
to the tensorial part.
The explicit tensorial expressions are presented  in the
Appendix,
using the graphical approach of Gaigalas {\it et al} (1985).
Graphical methods make it possible to reduce the number
of expressions from 42 to 6 for all
distributions presented in Table \ref{pasis}.
Such a joining up of several distributions is possible by
graphical means, because in the graphical technique
of Jucys and Bandzaitis (1977), as in the tensorial
products of operators of second quantization, the main
elements are the Clebsch-Gordan coefficients. Therefore
we may join up all the distributions having essentially
the same algebraic structure, although with different tensorial
products. The latter are represented by diagrams in which all
the pecularities of a tensorial product are seen, and the differences
of particular distributions are easily noticed. The use of other graphical
methods (see e.g. Yutsis  {\it et al } 1962 or Lindgren and Morrison 1982)
in joining up the distributions is complicated, since there the Wigner
coefficients play the main role, and these are not fully compatible
with the graphical transformations of the operators of second quantization
in coupled form.
Having classified the operators, we will now consider matrix
elements of these operators for arbitrary configurations.

\section{\bf Matrix elements between complex configurations}

Now, having the irreducible tensorial form of the operator
being considered, we are in a position to find their matrix elements
and recoupling matrices.
Suppose that we have a bra function with $u$ shells in $LS$ coupling:
% CFF This is confusing since now \alpha represents a quantum number but
%     in the next equation \aha becomes an index again.
\begin{equation}
\label{eq:ra}
\begin{array}[b]{c}
\psi _u^{bra}\left( LSM_LM_S\right) \equiv \\
\equiv (n_1l_1n_2l_2...n_ul_u\alpha _1L_1S_1Q_1M_{Q_1}\alpha
_2L_2S_2Q_2M_{Q_2}...\alpha _uL_uS_uQ_uM_{Q_u}{\cal A}LSM_LM_S|
\end{array}
\end{equation}
and a ket function:
\begin{equation}
\label{eq:rb}
\begin{array}[b]{c}
\psi _u^{ket}\left( L^{\prime }S^{\prime }M^{\prime}_LM^{\prime}_S\right) \equiv \\
\equiv |n_1l_1n_2l_2...n_ul_u\alpha _1^{\prime }L_1^{\prime }S_1^{\prime
}Q_1^{\prime }M_{Q_1}^{\prime }\alpha _2^{\prime }L_2^{\prime }S_2^{\prime
}Q_2^{\prime }M_{Q_2}^{\prime }...\alpha _u^{\prime }L_u^{\prime
}S_u^{\prime }Q_u^{\prime }M_{Q_u}^{\prime }{\cal A}^{\prime }L^{\prime }
S^{\prime}M^{\prime}_LM^{\prime}_S)
\end{array}
\end{equation}
where ${\cal A}$ stands for all intermediate quantum numbers, depending on the
order of coupling of momenta $L_iS_i$. Label $Q_i$ is the quasispin momentum of
the shell $n_il^{N_i}_{i}$, which is related to the seniority quantum
number $\nu_i $, namely, $%
Q_i=\left( 2l_i+1-\nu_i \right) /2$, and its projection,
$M_{Q_i}=\left( N_i-2l_i-1\right)
/2$. In (\ref{eq:ra}) and (\ref{eq:rb}) $\alpha _i$ denotes all additional
quantum numbers needed for the classification of the energy
levels of the relevant shell.

Using the Wigner-Eckart theorem in $LS$ space we shift from the
matrix element of any two-particle operator $G$
between functions (\ref{eq:ra}) and (\ref{eq:rb})
to the submatrix element
$(\psi _u^{bra}\left( LS\right) ||G||\psi _u^{ket}\left( L^{\prime
}S^{\prime }\right) )$ of this operator.

A general expression for the submatrix element of
any two-particle operator between functions (\ref{eq:ra}) and (\ref{eq:rb})
with $u$ open shells can be written as

\begin{equation}
\label{eq:mg}
\begin{array}[b]{c}
(\psi _u^{bra}\left( LS\right) ||G||\psi _u^{ket}\left( L^{\prime
}S^{\prime }\right) )= \\
=
\displaystyle {\sum_{n_il_i,n_jl_j,n_i^{\prime }l_i^{\prime },n_j^{\prime
}l_j^{\prime }}}\displaystyle {\sum_{\kappa _{12},\sigma _{12},\kappa
_{12}^{\prime },\sigma _{12}^{\prime }}}\sum \left( -1\right) ^\Delta \Theta
^{\prime }\left( n_i\lambda _i,n_j\lambda _j,n_i^{\prime }\lambda _i^{\prime
},n_j^{\prime }\lambda _j^{\prime },\Xi \right) \times \\ \times T\left(
n_i\lambda _i,n_j\lambda _j,n_i^{\prime }\lambda _i^{\prime },n_j^{\prime
}\lambda _j^{\prime },\Lambda ^{bra},\Lambda ^{ket},\Xi ,\Gamma \right)
\times \\
\times R\left( \lambda _i,\lambda _j,\lambda _i^{\prime },\lambda _j^{\prime
},\Lambda ^{bra},\Lambda ^{ket},\Gamma \right)
\end{array}
\end{equation}
where $\Lambda ^{bra}\equiv \left( L_iS_i,L_jS_j,L_i^{\prime }S_i^{\prime
},L_j^{\prime }S_j^{\prime }\right) ^{bra}$ is the array for the bra function
shells' terms, and similarly for $\Lambda ^{ket}$. This
expression is similar to Eq. (136) used by Grant (1988) in his
derivation. So, to calculate the
spin-angular part of a submatrix element, one has to compute:

\begin{enumerate}
\item  The recoupling matrix $R\left( \lambda _i,\lambda _j,\lambda
_i^{\prime },\lambda _j^{\prime },\Lambda ^{bra},\Lambda ^{ket},\Gamma
\right) $. This recoupling matrix accounts for the change in
going from matrix element $(\psi _u^{bra}\left( LS\right) ||
G||\psi _u^{ket}\left( L^{\prime }S^{\prime }\right) )$
, which has $u$ open
shells in the bra and ket functions, to the submatrix element $T\left(
n_i\lambda _i,n_j\lambda _j,n_i^{\prime }\lambda _i^{\prime },n_j^{\prime
}\lambda _j^{\prime },\Lambda ^{bra},\Lambda ^{ket},\Xi ,\Gamma \right) $,
which has only the shells being  acted upon by the two-particle operator in its bra and ket
functions.

\item
The submatrix element
$T\left( n_i\lambda _i,n_j\lambda _j,n_i^{\prime }\lambda _i^{\prime
},n_j^{\prime }\lambda _j^{\prime },\Lambda ^{bra},\Lambda ^{ket},\Xi
,\Gamma \right) $, which
denotes
the submatrix elements of operators of the types $A^{\left( kk'\right)
}\left( n\lambda ,\Xi \right) $, $B^{\left( kk'\right) }(n\lambda ,\Xi )$, $%
C^{\left( kk'\right) }(n\lambda ,\Xi )$, $D^{\left( ls\right) }$, $E^{\left(
kk'\right) }(n\lambda ,\Xi )$ (see (\ref{eq:mc})-(\ref{eq:mf})).
Here $\Gamma $ refers to the array of
coupling parameters
connecting the recoupling matrix \\ $R\left( \lambda _i,\lambda _j,\lambda
_i^{\prime },\lambda _j^{\prime },\Lambda ^{bra},\Lambda ^{ket},\Gamma
\right) $ to the submatrix element.

% CFF this does not make sense
%, because for
%any given distribution of electronic shells, upon which the second
%quantization operators are acting, the tensorial part of two-particle
%operator is expressed in terms of the operators
%(\ref{eq:mc})-(\ref{eq:mf}).
%presented earlier.

%$T\left( n_i\lambda _i,n_j\lambda _j,n_i^{\prime }\lambda _i^{\prime
%},n_j^{\prime }\lambda _j^{\prime },\Lambda ^{bra},\Lambda ^{ket},\Xi
%,\Gamma \right) $.

\item  Phase factor $\Delta $.

\item  $\Theta ^{\prime }\left( n_i\lambda _i,n_j\lambda _j,n_i^{\prime
}\lambda _i^{\prime },n_j^{\prime }\lambda _j^{\prime },\Xi \right) $, which
is proportional to the radial part and corresponds to one of \\
$\Theta \left(
n\lambda ,\Xi \right) $,...,$\Theta \left( n_\alpha \lambda _\alpha ,n_\beta
\lambda _\beta ,n_\gamma \lambda _\gamma ,n_\delta \lambda _\delta ,\Xi
\right) $. It consists of a submatrix element \\ $\left( n_i\lambda
_in_j\lambda _j||g^{\left( \kappa _1\kappa _2k,\sigma _1\sigma _2k\right)
}||n_i^{\prime }\lambda _i^{\prime }n_j^{\prime }\lambda _j^{\prime }\right)
$, and in some cases of simple factors and 3$nj$-coefficients. For
instance, for the distributions $\alpha \alpha \beta \beta $, $\gamma \gamma
\alpha \beta $, $\gamma \gamma \beta \alpha $, $\alpha \beta \gamma \gamma $%
, $\beta \alpha \gamma \gamma $, $\alpha \beta \gamma \delta $, $\beta
\alpha \delta \gamma $, $\alpha \beta \delta \gamma $, $\beta \alpha \gamma
\delta $, $\gamma \delta \alpha \beta $, $\delta \gamma \beta \alpha $, $%
\gamma \delta \beta \alpha $, $\delta \gamma \alpha \beta $ (see expressions
(\ref{eq:ac-a}), (\ref{eq:ac}) and notes on $\Theta ^{\prime }$ and $\tilde
\Theta $ in Appendix) it is:
\begin{equation}
\label{eq:mh}
\begin{array}[b]{c}
\Theta ^{\prime }\left( n_i\lambda _i,n_j\lambda _j,n_i^{\prime }\lambda
_i^{\prime },n_j^{\prime }\lambda _j^{\prime },\Xi \right) =\left( -1\right)
^t\tilde \Theta \left( n_i\lambda _i,n_j\lambda _j,n_i^{\prime }\lambda
_i^{\prime },n_j^{\prime }\lambda _j^{\prime },\Xi \right) = \\
=\frac 12\left( -1\right) ^{k-p+t+1}\left( n_i\lambda _in_j\lambda
_j||g^{\left( \kappa _1\kappa _2k,\sigma _1\sigma _2k\right) }||n_i^{\prime
}\lambda _i^{\prime }n_j^{\prime }\lambda _j^{\prime }\right) \times  \\
\times \left[ \kappa _{12},\sigma _{12},\kappa _{12}^{\prime },\sigma
_{12}^{\prime }\right] ^{1/2}\left\{
\begin{array}{ccc}
l_i & l_i^{\prime } & \kappa _1 \\
l_j & l_j^{\prime } & \kappa _2 \\
\kappa _{12} & \kappa _{12}^{\prime } & k
\end{array}
\right\} \left\{
\begin{array}{ccc}
s & s & \sigma _1 \\
s & s & \sigma _2 \\
\sigma _{12} & \sigma _{12}^{\prime } & k
\end{array}
\right\} ,
\end{array}
\end{equation}
where the integer $t$ determining the phase depends upon the configuration
states involved. Rules for its determination are given in the Appendix.

\end{enumerate}

The calculation of $\Theta ^{\prime }\left( n_i\lambda _i,n_j\lambda
_j,n_i^{\prime }\lambda _i^{\prime },n_j^{\prime }\lambda _j^{\prime },\Xi
\right) $ is straightforward (from an angular momentum point of
view) and depends on the radial form of
the operator.  In the next sections we will describe expressions
for the recoupling matrix, the submatrix elements, and the phase
factor, respectively.

\section{\bf Recoupling matrices}

In this section we present the expressions for the recoupling matrices
$$
R\left( \lambda _i,\lambda _j,\lambda _i^{\prime },\lambda _j^{\prime
},\Lambda ^{bra},\Lambda ^{ket},\Gamma \right) .
$$
These matrices  may be treated in the orbital $l$
and spin $s$ spaces separately. That is,

\begin{equation}
\label{eq:mi}
\begin{array}[b]{c}
R\left( \lambda _i,\lambda _j,\lambda _i^{\prime },\lambda _j^{\prime
},\Lambda ^{bra},\Lambda ^{ket},\Gamma \right) = \\
=R\left( l_i,l_j,l_i^{\prime },l_j^{\prime },\Lambda _l^{bra},\Lambda
_l^{ket},\Gamma _l\right) R\left( s,s,s,s,\Lambda _s^{bra},\Lambda
_s^{ket},\Gamma _s\right)
\end{array}
\end{equation}
where $\Lambda _l^{bra}\equiv \left( L_i,L_j,L_i^{\prime },L_j^{\prime
}\right) ^{bra}$ and $\Lambda _s^{bra}\equiv \left( S_i,S_j,S_i^{\prime
},S_j^{\prime }\right) ^{bra}$. Therefore for simplicity we present only the
expressions in $l$ space. The recoupling matrices in $s$ space are easily
obtained from analogous expressions in $l$ space by making corresponding
substitutions $l_1$, $l_2$,...,$l_u$ $\longrightarrow $ $s$; $%
L_1\longrightarrow S_1$, $L_2\longrightarrow S_2$;...; $L_{12}%
\longrightarrow S_{12}$,..., $L_{123..u-1}\longrightarrow S_{123...u-1}$; $%
L\longrightarrow S$, $L^{\prime }\longrightarrow S^{\prime }$. Also, the
analytical expressions for recoupling matrices presented in this section are
valid in the case of $jj$-coupling.

As we have mentioned earlier, there are four classes as defined
by Eq.'s (\ref{eq:mc})-(\ref{eq:mf}).
We will consider each class separately.
All the expressions presented below are obtained by using the approach of
angular momentum theory described by Jucys and Bandzaitis (1977).

\subsection{One interacting shell}

Let us assume that the operators of second quantization act upon shell $%
a $ as in distribution 1 of Table \ref{pasis}, where $a\equiv
\alpha $. Then the recoupling matrix has the expression:
\begin{equation}
\label{eq:rc}
\begin{array}{c}
R\left( l_a,L_a,k\right) = \\
=\left[ L_a\right] ^{-1/2}\delta \left( L_1,L_1^{\prime }\right) ...\delta
\left( L_{a-1},L_{a-1}^{\prime }\right) \delta \left(
L_{a+1},L_{a+1}^{\prime }\right) ...\delta \left( L_u,L_u^{\prime }\right)
\times \\
\\
\times \left\{
\begin{array}{ll}
\delta \left( L_1,L_1^{\prime },k\right) ; &\mbox{ for } u=1 \\ \ \\
C_1; & \mbox{ for } u=2 \\ \ \\
C_1C_2\left( k,a+1,u-1\right) C_3; & \mbox{ for }a<3,u>2 \\\ \\
\begin{array}{c}
\delta \left( L_{12},L_{12}^{\prime }\right) ...\delta \left(
L_{12...a-1},L_{12...a-1}^{\prime }\right) \times \\
\times C_1C_2\left( k,a+1,u-1\right) C_3;
\end{array}
& \mbox{ for } a>3,a\neq u,u>2 \\ \ \\
\delta \left( L_{12},L_{12}^{\prime }\right) ...\delta \left(
L_{12...a-1},L_{12...a-1}^{\prime }\right) C_3; & \mbox{ for } a=u,u>2
\end{array}
\right.
\end{array}
\end{equation}
In the above,
the notation $\delta \left( L_1,L_1^{\prime }, k\right)$
means the triangular condition $\mid L_1-L_1^{\prime } \mid \leq k \leq
L_1+L_1^{\prime }$ and
\begin{equation}
\label{eq:rd}C_1=\left( -1\right) ^\varphi \left[ L_a,T^{\prime }\right]
^{1/2}\left\{
\begin{array}{ccc}
k & L_a^{\prime } & L_a \\
J & T & T^{\prime }
\end{array}
\right\} ,
\end{equation}
where the values of parameters $\varphi$, $J$, $T$ and $T^{\prime}$ present in
expression (\ref{eq:rd}) are given in Table \ref{Ca}.
\begin{table}
\begin{center}
%\caption{Parameters for equation (\ref{eq:rd}) }
\caption{Parameters for equation (15)}
\label{Ca}
\begin{tabular}{|l|l|l|l|l|l|} \hline
          &       &  &  &  & \\
 $u$      & $a$   & $\varphi$  & $J$ & $T$ & $T^{\prime} $ \\
          &       &  &  &  &  \\ \hline  \hline
          &       &  &  &  & \\
 2        &1&$L_1+2L_1^{\prime }-L_2-L^{\prime }+k$&$L_{2,}$&$L$&$L^{\prime }$\\
 2        &2&$L_1+L+L_2^{\prime }+k$               &$L_1$   &$L$&$L^{\prime }$\\
 $u\neq 2$&1&$L_1+2L_1^{\prime }-L_2-L_{12}^{\prime }+k$&$L_{2,}$&$L_{12}$&
                                                            $L_{12}^{\prime }$\\
 $u\neq 2$&2&$L_1+L_{12}+L_2^{\prime }+k$   &$L_1$&$L_{12}$&$L_{12}^{\prime }$\\
 $u\neq 2$&$a>2$&$L_{12...a-1}+L_{12...a}+L_a^{\prime }+k$&$L_{12...a-1}$&
                   $L_{12...a}$&$L_{12...a}^{\prime }$\\
          &       &  &  &  &  \\ \hline
\end{tabular}
\end{center}
\end{table}
The remaining two coefficients are
\begin{equation}
\label{eq:re}
\begin{array}[b]{c}
C_2\left( k,k_{\min },k_{\max }\right) =
\displaystyle {\stackrel{k_{\max }}{\prod_{i=k_{\min }}}}\left( -1\right)
^{k+L_i+L_{12...i-1}+L_{12...i}^{\prime }}
\left[ L_{12...i-1},L_{12...i}^{\prime
}\right] ^{1/2}\times \\ \times \left\{
\begin{array}{ccc}
k & L_{12...i-1}^{\prime } & L_{12...i-1} \\
L_i & L_{12...i} & L_{12...i}^{\prime }
\end{array}
\right\} ;
\end{array}
\end{equation}
and
\begin{equation}
\label{eq:rf}C_3=\left( -1\right) ^\varphi \left[ J,T^{\prime }\right]
^{1/2}\left\{
\begin{array}{ccc}
k & J^{\prime } & J \\
j & T & T^{\prime }
\end{array}
\right\} ;
\end{equation}
where the parameters $\varphi$, $j$, $J$, $J^{\prime}$, $T$ and $T^{\prime}$
are given in Table \ref{Cc}.

\begin{table}
\begin{center}
%\caption{Parameters for equation (\ref{eq:rf}) }
\caption{Parameters for equation (17)}
\label{Cc}
\begin{tabular}{|l|l|l|l|l|l|l|} \hline
         &       &  &  &  & & \\
 $u$     & $\varphi$ & $j$ & $J$ & $J^{\prime}$ &$T$ & $T^{\prime} $ \\
         &       &  &  &  & &  \\ \hline  \hline
         &       &  &  &  & & \\
$u\neq a$&$k+L_u+L_{12...u-1}+L^{\prime }$&$L_u$&$L_{12...u-1}$&
          $L_{12...u-1}^{\prime }$        &$L$  &$L^{\prime }$ \\
$a$      &$k-L_{12...u-1}+2L_u+L_u^{\prime }-L$&$L_{12...u-1}$&$L_u$&
          $L_u^{\prime }$&$L$&$L^{\prime }$ \\
         &       &  &  &  & &  \\ \hline
\end{tabular}
\end{center}
\end{table}

When the total rank $k=0$, the recoupling matrix becomes simply
\begin{equation}
\label{eq:rg}
\begin{array}[b]{c}
R\left( l_a,L_a,0\right) =\delta \left( L_1,L_1^{\prime }\right) \delta
\left( L_2,L_2^{\prime }\right) \delta \left( L_{12},L_{12}^{\prime }\right)
...\delta \left( L_{a-1},L_{a-1}^{\prime }\right) \times \\
\times \delta \left( L_{12...a-1},L_{12...a-1}^{\prime }\right) \delta
\left( L_a,L_a^{\prime }\right) \delta \left( L_{12...a},L_{12...a}^{\prime
}\right) \delta \left( L_{a+1},L_{a+1}^{\prime }\right) \times \\
\times \delta \left( L_{12...a+1},L_{12...a+1}^{\prime }\right) ...\delta
\left( L_u,L_u^{\prime }\right) \delta \left( L,L^{\prime }\right) .
\end{array}
\end{equation}
Expression (\ref{eq:rg}) is equivalent to (13.60) of Cowan (1981).

\subsection{Two interacting shells}

In this case let us assume that the operators of second quantization act
upon the shells $a$ and $b$ (distributions 2-10 in Table \ref{pasis}, where
for distributions 2-5 $a\equiv \alpha $, $b\equiv \beta $ and for
others (6-10) $a=\min \{\alpha ,\beta \}$, $b=\max \{\alpha ,\beta
\} $). Then

\begin{equation}
\label{eq:rh}
\begin{array}[b]{c}
R\left( l_a,L_a,l_b,L_b,\kappa _{12},\kappa _{12}^{\prime },k\right) =\left(
-1\right) ^\zeta \left[ L_a,L_b\right] ^{-1/2}\delta \left( L_1,L_1^{\prime
}\right) ...\delta \left( L_{a-1},L_{a-1}^{\prime }\right) \times \\
\times \delta \left( L_{a+1},L_{a+1}^{\prime }\right) ...\delta \left(
L_{b-1},L_{b-1}^{\prime }\right) \delta \left( L_{b+1},L_{b+1}^{\prime
}\right) ...\delta \left( L_u,L_u^{\prime }\right) \times \\
\\
\times \left\{
\begin{array}{ll}
C_4\left( K_{12},K_{12}^{\prime },k,1\right) C_2\left( k,3,u-1\right) C_3; &
\mbox{ for } a=1,b=2 \\ \ \\
\begin{array}{l}
C_1C_2\left( K_{12},a+1,b-1\right) C_4\left( K_{12},K_{12}^{\prime
},k,1\right) \times \\
\times C_2\left( k,b+1,u-1\right) C_3;
\end{array}
& \mbox{ for } a<3,b>2,b\neq u \\ \ \\
C_1C_2\left( K_{12},a+1,b-1\right) C_4\left( K_{12},K_{12}^{\prime
},k,1\right) ; &  \mbox{ for }a<3,b=u \\ \ \\
\begin{array}{l}
\delta \left( L_{12},L_{12}^{\prime }\right) ...\delta \left(
L_{12...a-1},L_{12...a-1}^{\prime }\right) C_1\times \\
\times C_2\left( K_{12},a+1,b-1\right) C_4\left( K_{12},K_{12}^{\prime
},k,1\right) \times \\
\times C_2\left( k,b+1,u-1\right) C_3;
\end{array}
& \mbox{ for } a\geq 3,b>2,b\neq u \\ \ \\
\begin{array}{l}
\delta \left( L_{12},L_{12}^{\prime }\right) ...\delta \left(
L_{12...a-1},L_{12...a-1}^{\prime }\right) C_1\times \\
\times C_2\left( K_{12},a+1,b-1\right) C_4\left( K_{12},K_{12}^{\prime
},k,1\right) ;
\end{array}
& \mbox{ for }a\geq 3,b=u
\end{array}
\right.
\end{array}
\end{equation}
where
\begin{equation}
\label{eq:ph-a}\zeta =\left\{
\begin{array}{cc}
0 & \mbox{ for } \alpha <\beta , \\
\kappa _{12}+\kappa _{12}^{\prime }-k & \mbox{ for } \alpha >\beta ,
\end{array}
\right.
\end{equation}
and
\begin{equation}
\label{eq:rk}C_4\left( k_1,k_2,k,P\right) =\left[ J_1,J_2,J_3^{\prime
},k\right] ^{1/2}\left\{
\begin{array}{ccc}
J_1^{\prime } & k_1 & J_1 \\
J_2^{\prime } & k_2 & J_2 \\
J_3^{\prime } & k & J_3
\end{array}
\right\} .
\end{equation}

The values of parameters $J_1$, $J_1^{\prime}$, $J_2$, $J_2^{\prime}$, $J_3$
and $J_3^{\prime}$ present in the expression (\ref{eq:rk}) must be taken
from Table \ref{Cd}.
\begin{table}
\begin{center}
%\caption{Parameters for equation (\ref{eq:rk}) }
\caption{Parameters for equation (21)}
\label{Cd}
  \begin{tabular}{|l|lll|l|l|l|l|l|l|} \hline
   &       &  &  &  & & & & & \\
$P$&$a$&$b$&$u$&$J_1$&$J_1^{\prime}$&$J_2$&$J_2^{\prime}$&$J_3$&$J_3^{\prime}$ \\
   &       &  &  &  & & & & & \\ \hline \hline
   &       &  &  &  & & & & & \\
$1$&$1$&$2$&$u\ne b$&$L_1$&$L_1^{\prime}$&$L_2$&$L_2^{\prime}$&$L_{12}$&$L_{12}$ \\
$1$&$1$&$2$&$b$    &$L_1$&$L_1^{\prime }$&$L_2$&$L_2^{\prime }$&$L$&$L$ \\
$1$&$a\neq 1$&$b\neq 2$&$b$& $L_{1...u-1}$&$L_{1...u-1}^{\prime}$&$L_u$&
                $L_u^{\prime}$&$L$&$L^{\prime}$ \\
$1$&\multicolumn{3}{c|}{in all other cases}&$L_{1...b-1}$&$L_{1...b-1}^{\prime }$&$L_b$&
                $L_b^{\prime }$&$L_{1...b}$&$L_{1...b}^{\prime }$ \\
   &       &  &  &  & & & & & \\ \hline
   &       &  &  &  & & & & & \\
$2$&\multicolumn{3}{c|}{in all cases} &$L_{1...c-1}$&$L_{1...c-1}^{\prime }$&$L_c$&
                $L_c^{\prime }$&$L_{1...c}$&$L_{1...c}^{\prime }$ \\
   &       &  &  &  & & & & & \\ \hline
\end{tabular}
\end{center}
\end{table}
For the case $\alpha <\beta $ in Eq. (\ref{eq:rh}) $K_{12}=\kappa _{12}$, $
K_{12}^{\prime }=\kappa _{12}^{\prime }$ and when $\alpha >\beta ,$ then $%
K_{12}=\kappa _{12}^{\prime }$, $K_{12}^{\prime }=\kappa _{12}$.
When the total rank $k=0$, and $\kappa _{12}=\kappa _{12}^{\prime }=k$, the
recoupling matrix has the form:
\begin{equation}
\label{eq:rl}
\begin{array}{c}
R\left( l_a,L_a,l_b,L_b,k,k,0\right) =\left[ L_a,L_b^{\prime },k\right]
^{-1/2}\delta \left( L_1,L_1^{\prime }\right) ...\delta \left(
L_{a-1},L_{a-1}^{\prime }\right) \times \\
\times \delta \left( L_{a+1},L_{a+1}^{\prime }\right) ...\delta \left(
L_{b-1},L_{b-1}^{\prime }\right) \delta \left( L_{b+1},L_{b+1}^{\prime
}\right) ...\delta \left( L_u,L_u^{\prime }\right) \times \\
\times \delta \left( L_{12},L_{12}^{\prime }\right) ...\delta \left(
L_{12...a-1},L_{12...a-1}^{\prime }\right) \delta \left(
L_{12...b},L_{12...b}^{\prime }\right) ...\delta \left( L,L^{\prime }\right)
\times \\
\\
\times \left\{
\begin{array}{ll}
C_5\left( 1\right) ; & \mbox{ for }a=1,b=2 \\ \ \\
C_1C_2\left( k,a+1,b-1\right) C_5\left( 1\right) ; & \mbox{ for
}a<3 \\ \ \\
\begin{array}{l}
\delta \left( L_{12},L_{12}^{\prime }\right) ...\delta \left(
L_{12...a-1},L_{12...a-1}^{\prime }\right) \times \\
\times C_1C_2\left( k,a+1,b-1\right) C_5\left( 1\right) ;
\end{array}
& \mbox{ for } a\geq 3
\end{array}
\right. ,
\end{array}
\end{equation}
where
\begin{equation}
\label{eq:rm}C_5\left( P\right) =\left( -1\right) ^{k+L_b+J_1^{\prime
}+J_2}\left[ J_1,L_b^{\prime }\right] ^{1/2}\left\{
\begin{array}{ccc}
k & L_b^{\prime } & L_b \\
J_2 & J_1 & J_1^{\prime }
\end{array}
\right\} .
\end{equation}
The values of parameters $J_1$, $J_1^{\prime }$ and $J_2$ present in the
expression (\ref{eq:rm}) must be taken from Table \ref{Ce}.

\begin{table}
\begin{center}
%\caption{Parameters for equation (\ref{eq:rm}) }
\caption{Parameters for equation (23)}
\label{Ce}
\begin{tabular}{|l|l|l|l|l|} \hline
   &       &  &  &  \\
$P$&Case&$J_1$&$J_1^{\prime }$&$J_2$  \\
   &       &  &  &  \\  \hline \hline
   &       &  &  &  \\
1  &$a=1$ and $b=2$&$L_a$&$L_a^{\prime}$&$L_{12...b}$ \\
1  &$b\neq u$            &$L_{1...b-1}$&$L_{1...b-1}^{\prime }$&$L_{12...b}$ \\
1  &$b=u$                &$L_{1...b-1}$&$L_{1...b-1}^{\prime }$&$L$ \\
   &       &  &  &  \\  \hline
   &       &  &  &  \\
2  &$c\neq u$            &$L_{1...c-1}$&$L_{1...c-1}^{\prime }$&$L_{12...c}$ \\
2  &$c=u$                &$L_{1...c-1}$&$L_{1...c-1}^{\prime }$&$L$ \\
   &       &  &  &  \\  \hline
   &       &  &  &  \\
3  &$d\neq u$            &$L_{1...d-1}$&$L_{1...d-1}^{\prime }$&$L_{12...d}$ \\
3  &$d=u$                &$L_{1...d-1}$&$L_{1...d-1}^{\prime }$&$L$ \\
   &       &  &  &  \\  \hline
\end{tabular}
\end{center}
\end{table}

Formula (\ref{eq:rl}) has no analogue in Cowan (1981). Our expressions for
the recoupling matrix do not depend on coefficients of fractional parentage and
have no intermediate summations. Therefore they will be very convenient for
practical calculations.

\subsection{Three interacting shells}

When the operators of second quantization act upon three shells $a$, $b$ and
$c$ (distributions 11-18 in the Table \ref{pasis}), we have:

\begin{equation}
\label{eq:rn}
\begin{array}{c}
R\left( l_a,L_a,l_b,L_b,l_c,L_c,k_1,k_2,\kappa _{12},\kappa _{12}^{\prime
},k\right) = \\
=\left[ L_a,L_b,L_c\right] ^{-1/2}\delta \left( L_1,L_1^{\prime }\right)
...\delta \left( L_{a-1},L_{a-1}^{\prime }\right) \times \\
\times \delta \left( L_{a+1},L_{a+1}^{\prime }\right) ...\delta \left(
L_{b-1},L_{b-1}^{\prime }\right) \delta \left( L_{b+1},L_{b+1}^{\prime
}\right) ...\delta \left( L_{c-1},L_{c-1}^{\prime }\right) \times \\
\times \delta \left( L_{c+1},L_{c+1}^{\prime }\right) ...\delta \left(
L_u,L_u^{\prime }\right)
\displaystyle {\sum_{j_{12}}}\left( -1\right) ^\zeta C_6\times \\  \\
\times \left\{
\begin{array}{ll}
\begin{array}{l}
C_4\left( K_1,K_2,j_{12},1\right) C_2\left( j_{12},3,c-1\right) \times \\
\times C_4\left( j_{12},K_3,k,2\right) C_2\left( k,c+1,u-1\right) C_3;
\end{array}
&  \mbox{ for }a=1,b=2 \\ \ \\
\begin{array}{l}
C_1C_2\left( K_1,a+1,b-1\right) C_4\left( K_1,K_2,j_{12},1\right) \times \\
\times C_2\left( j_{12},b+1,c-1\right) C_4\left( j_{12},K_3,k,2\right)
\times \\
\times C_2\left( k,c+1,u-1\right) C_3;
\end{array}
&  \mbox{ for }a<3 \\ \ \\
\begin{array}{l}
\delta \left( L_{12},L_{12}^{\prime }\right) ...\delta \left(
L_{12...a-1},L_{12...a-1}^{\prime }\right) C_1\times \\
\times C_2\left( K_1,a+1,b-1\right) C_4\left( K_1,K_2,j_{12},1\right) \times
\\
\times C_2\left( j_{12},b+1,c-1\right) C_4\left( j_{12},K_3,k,2\right)
\times \\
\times C_2\left( k,c+1,u-1\right) C_3;
\end{array}
& \mbox{ for } a\geq 3,
\end{array}
\right.
\end{array}
\end{equation}
where parameters $a$, $b$, $c$, $\zeta$, $K_1$, $K_2$, $K_3$ and
coefficient $C_6$ are given in Table \ref{Rc}.
\begin{table}
\begin{center}
%\caption{Parameters for equation (\ref{eq:rn}) }
\caption{Parameters for equation (24)}
\label{Rc}
\begin{tabular}{|l|l|l|l|l|l|l|l|l|} \hline
   &       &  &  &  & & & & \\
Case&$a$&$b$&$c$&$\zeta$&$K_1$&$K_2$&$K_3$&$C_6$ \\
   &       &  &  &  & & & & \\  \hline \hline
   &       &  &  &  & & & & \\
$\alpha <\beta <\gamma $&$\alpha$&$\beta$&$\gamma$&$0$&$k_1$&$k_2$
&$\kappa _{12}^{\prime }$&$\delta \left( j_{12},\kappa _{12}\right)$ \\
$\beta <\alpha <\gamma$&$\beta$&$\alpha$&$\gamma$&$k_1+k_2-\kappa _{12}$&$k_2$&
$k_1$&$\kappa_{12}^{\prime }$&$\delta \left( j_{12},\kappa _{12}\right)$ \\
$\beta <\gamma <\alpha$&$\beta$&$\gamma$&$\alpha$&$0$&$k_2$&
$\kappa_{12}^{\prime}$&$k_1$&
$C_6^{\prime}\left(\kappa_{12}^{\prime},k_2,j_{12},k_1,k,\kappa_{12}\right)$ \\
$\alpha <\gamma <\beta$&$\alpha$&$\gamma$&$\beta$&$k_1+k_2-\kappa _{12}$&
$k_1$&$\kappa _{12}^{\prime }$&$k_2$&
$C_6^{\prime}\left(\kappa_{12}^{\prime},k_1,j_{12},k_2,k,\kappa_{12}\right)$ \\
$\gamma <\alpha <\beta$&$\gamma$&$\alpha$&$\beta$&
$2k_1+k_2-\kappa_{12}+$&$\kappa_{12}^{\prime }$&$k_1$&$k_2$&
$C_6^{\prime}\left(\kappa_{12}^{\prime},k_1,j_{12},k_2,k,\kappa_{12}\right)$ \\
$$&$$&$$&$$&
$+\kappa^{\prime}{}_{12}-j_{12}$&$$&$$&$$&$$ \\
$\gamma <\beta <\alpha$&$\gamma$&$\beta$&$\alpha$&$k_1+k_2-\kappa _{12}$&
$\kappa _{12}^{\prime }$&$k_2$&$k_1$&
$C_6^{\prime}\left(\kappa_{12}^{\prime},k_2,j_{12},k_1,k,\kappa_{12}\right)$ \\
   &       &  &  &  & & & & \\  \hline
\end{tabular}
\end{center}
\end{table}
The coefficient

$C_6^{\prime }\left( k_1,k_2,k_3,k_4,k_5,k_6\right) $ is
\begin{equation}
\label{eq:coe-s}C_6^{\prime }\left( k_1,k_2,k_3,k_4,k_5,k_6\right) =\left(
-1\right) ^{k_1+k_2-k_3+2k_5}\left[ k_3,k_6\right] ^{1/2}\left\{
\begin{array}{ccc}
k_1 & k_2 & k_3 \\
k_4 & k_5 & k_6
\end{array}
\right\} .
\end{equation}
From (\ref{eq:me}) we have that in expressions (\ref{eq:rn})
and (\ref{eq:coe-s}) the ranks $k_1=l_\alpha $, $k_2=l_\beta $.

When the total rank $k=0$, and $\kappa _{12}=\kappa _{12}^{\prime }=k$, the
recoupling matrix has the form:
\begin{equation}
\label{eq:ro}
\begin{array}[b]{c}
R\left( l_a,L_a,l_b,L_b,l_c,L_c,k_1,k_2,k,k,0\right) =\left( -1\right)
^\zeta \left[ L_a,L_b,L_c^{\prime },K_3\right] ^{-1/2} \times \\
\times \delta \left( L_1,L_1^{\prime }\right) ...\delta \left(
L_{a-1},L_{a-1}^{\prime }\right) \delta \left( L_{a+1},L_{a+1}^{\prime
}\right) ...\delta \left( L_{b-1},L_{b-1}^{\prime }\right) \times \\
\times \delta \left( L_{b+1},L_{b+1}^{\prime }\right) ...\delta \left(
L_{c-1},L_{c-1}^{\prime }\right) \delta \left( L_{c+1},L_{c+1}^{\prime
}\right) ...\delta \left( L_u,L_u^{\prime }\right) \times \\
\times \delta \left( L_{12},L_{12}^{\prime }\right) ...\delta \left(
L_{12...a-1},L_{12...a-1}^{\prime }\right) \delta \left(
L_{12...c},L_{12...c}^{\prime }\right) ...\delta \left( L,L^{\prime }\right)
\times \\
\\
\times \left\{
\begin{array}{ll}
C_4\left( K_1,K_2,K_3,1\right) C_2\left( K_3,b+1,c-1\right) C_5\left(
2\right) ; & \mbox{ for } a=1,b=2 \\ \ \\
\begin{array}{l}
C_1C_2\left( K_1,a+1,b-1\right) C_4\left( K_1,K_2,K_3,1\right) \times \\
\times C_2\left( K_3,b+1,c-1\right) C_5\left( 2\right) ;
\end{array}
& \mbox{ for } a<3 \\ \ \\
\begin{array}{l}
\delta \left( L_{12},L_{12}^{\prime }\right) ...\delta \left(
L_{12...a-1},L_{12...a-1}^{\prime }\right) C_1\times \\
\times C_2\left( K_1,a+1,b-1\right) C_4\left( K_1,K_2,K_3,1\right) \times \\
\times C_2\left( K_3,b+1,c-1\right) C_5\left( 2\right) ;
\end{array}
& \mbox{ for } a\geq 3
\end{array}
\right.
\end{array}
\end{equation}
where the parameters $\zeta$, $K_1$, $K_2$, $K_3$ values
are given in Table \ref{Rcb}.

\begin{table}
\begin{center}
%\caption{Parameters for equation (\ref{eq:ro}) }
\caption{Parameters for equation (26)}
\label{Rcb}
  \begin{tabular}{|l|l|l|l|l|} \hline
   &       &  &  &   \\
Case&$\zeta$&$K_1$&$K_2$&$K_3$ \\
   &       &  &  &   \\  \hline \hline
   &       &  &  &   \\
$\alpha < \beta < \gamma$&$0$&$k_1$&$k_2$&$k$ \\
$\beta < \alpha < \gamma$&$k_1+k_2-k$&$k_2$&$k_1$&$k$ \\
$\beta < \gamma < \alpha$&$2k_1$&$k_2$&$k$&$k_1$ \\
$\alpha < \gamma < \beta$&$k_1-k_2-k$&$k_1$&$k$&$k_2$ \\
$\gamma < \alpha < \beta$&$2k$&$k$&$k_1$&$k_2$ \\
$\gamma < \beta < \alpha$&$k_1+k_2+k$&$k$&$k_2$&$k_1$ \\
   &       &  &  &   \\  \hline
\end{tabular}
\end{center}
\end{table}

The recoupling matrix for three interacting shells (\ref{eq:ro}) has
the same
advantages as the equivalent quantity, Eq. (\ref{eq:rl}),
for two shells.

\subsection{Four interacting shells}

When the operators of second quantization act upon four shells $a$, $b$, $c$
and $d$ (distributions 19-42 in the Table \ref{pasis}), we have:

\begin{equation}
\label{eq:keturi}
\begin{array}[b]{c}
R\left( l_a,L_a,l_b,L_b,l_c,L_c,l_d,L_d,k_1,k_2,\kappa _{12},k_3,k_4,\kappa
_{12}^{\prime },k\right) = \\
=\left[ L_a,L_b,L_c,L_d\right] ^{-1/2}\delta \left( L_1,L_1^{\prime }\right)
...\delta \left( L_{a-1},L_{a-1}^{\prime }\right) \times \\
\times \delta \left( L_{a+1},L_{a+1}^{\prime }\right) ...\delta \left(
L_{b-1},L_{b-1}^{\prime }\right) \delta \left( L_{b+1},L_{b+1}^{\prime
}\right) ...\delta \left( L_{c-1},L_{c-1}^{\prime }\right) \times \\
\times \delta \left( L_{c+1},L_{c+1}^{\prime }\right) ...\delta \left(
L_{d-1},L_{d-1}^{\prime }\right) \delta \left( L_{d+1},L_{d+1}^{\prime
}\right) ...\delta \left( L_u,L_u^{\prime }\right) \times \\
\\
\times \left\{
\begin{array}{ll}
\begin{array}{l}
C_4\left( k_1,k_2,\kappa _{12},1\right) C_2\left( \kappa _{12},3,c-1\right)
C_7\left( c,d\right) \times \\
\times C_2\left( k,d+1,u-1\right) C_3;
\end{array}
&  \mbox{ for }a=1,b=2 \\ \ \\
\begin{array}{l}
C_1C_2\left( k_1,a+1,b-1\right) C_4\left( k_1,k_2,\kappa _{12},1\right)
\times \\
\times C_2\left( \kappa _{12},b+1,c-1\right) C_7\left( c,d\right) \times \\
\times C_2\left( k,d+1,u-1\right) C_3;
\end{array}
& \mbox{ for } a<3 \\ \ \\
\begin{array}{l}
\delta \left( L_{12},L_{12}^{\prime }\right) ...\delta \left(
L_{12...a-1},L_{12...a-1}^{\prime }\right) C_1\times \\
\times C_2\left( k_1,a+1,b-1\right) C_4\left( k_1,k_2,\kappa _{12},1\right)
\times \\
\times C_2\left( \kappa _{12},b+1,c-1\right) C_7\left( c,d\right) \times \\
\times C_2\left( k,d+1,u-1\right) C_3;
\end{array}
& \mbox{ for } a\geq 3,
\end{array}
\right.
\end{array}
\end{equation}
where
\begin{equation}
\label{eq:r-a}
\begin{array}[b]{c}
C_7\left( k_{\min },k_{\max }\right) = \\
\\
=\left\{
\begin{array}{ll}
\displaystyle {\sum_I}C_8\left( I\right) C_{10}\left( I\right) , &
\mbox{ for }k_{\max }-k_{\min }=1 \\ \ \\
\displaystyle {\sum_{I_1}}\displaystyle {\sum_{I_2}}C_8\left( I_1\right)
C_9\left( I_1,I_2,k_{\min }+1\right) C_{10}\left( I_2\right) , &
\mbox{ for }k_{\max }-k_{\min }=2 \\ \ \\
\displaystyle {\sum_{I_{1}}}\displaystyle {\sum_{I_{2}}}%
C_8\left( I_{1}\right)
C_{11}\left(I_{1}, I_{2}\right)
C_{10}\left( I_{2}\right);
& \mbox{ for } k_{\max }-k_{\min }<2
\end{array}
\right.
\end{array}
\end{equation}

\begin{equation}
\label{eq:r-b}
\begin{array}[b]{c}
C_8\left( I\right) =\left( -1\right)
^{\kappa _{12}+L_{12...c}^{\prime }-I}\left[
L_c,I,L_{12...c-1},L_{12...c}^{\prime }\right] ^{1/2}\times \\
\times \left\{
\begin{array}{ccc}
k_3 & L_c^{\prime } & L_c \\
L_{12...c-1} & L_{12...c} & I
\end{array}
\right\} \left\{
\begin{array}{ccc}
L_{12...c-1}^{\prime } & \kappa _{12} & L_{12...c-1} \\
I & L_c^{\prime } & L_{12...c}^{\prime }
\end{array}
\right\} ,
\end{array}
\end{equation}

\begin{equation}
\label{eq:r-c}
\begin{array}[b]{c}
C_9\left( I_1,I_2,i\right) =\left( -1\right)
^{2(I_1+L_i)+L_{12...i}+L_{12...i}^{\prime }+k_3+\kappa _{12}}\left[
L_{12...i-1},I_1,I_2,L_{12...i}^{\prime }\right] ^{1/2}\times \\
\times \left\{
\begin{array}{ccc}
L_{12...i-1} & I_1 & k_3 \\
I_2 & L_{12...i} & L_i
\end{array}
\right\} \left\{
\begin{array}{ccc}
L_{12...i-1}^{\prime } & I_1 & \kappa _{12} \\
I_2 & L_{12...i}^{\prime } & L_i
\end{array}
\right\} ,
\end{array}
\end{equation}

\begin{equation}
\label{eq:r-d}
\begin{array}[b]{c}
C_{10}\left( I\right) =\left( -1\right) ^{2\left( I+k_3\right) +k_4+\kappa
_{12}+\kappa _{12}^{\prime }+k+L_{12...d}+L_{12...d}^{\prime
}+L_d+L_d^{\prime }+L_{12...d-1}^{\prime }}\times \\
\times \left[ \kappa _{12},\kappa _{12}^{\prime },L_d,I,L_{12...d}^{\prime
},L_{12...d-1}\right] ^{1/2}
\displaystyle {\sum_x}\left( -1\right) ^x\left[ x\right] \times \\ \times
\left\{
\begin{array}{ccc}
I & \kappa _{12}^{\prime } & x \\
k & L_{12...d-1}^{\prime } & \kappa _{12}
\end{array}
\right\} \left\{
\begin{array}{ccc}
I & \kappa _{12}^{\prime } & x \\
k_4 & L_{12...d-1} & k_3
\end{array}
\right\} \times \\
\times \left\{
\begin{array}{ccc}
L_{12...d-1} & k_4 & x \\
L_d^{\prime } & L_{12...d} & L_d
\end{array}
\right\} \left\{
\begin{array}{ccc}
L_{12...d-1}^{\prime } & k & x \\
L_{12...d} & L_d^{\prime } & L_{12...d}^{\prime }
\end{array}
\right\} .
\end{array}
\end{equation}

\begin{equation}
\label{eq:r-van}
\begin{array}[b]{c}
C_{11}\left( I_1,I_2\right) =\left( -1\right) ^{I_1-I_2
+L_{12...c}^{\prime}-L_{12...d-1}^{\prime}}\left[ I_1,I_2\right] ^{1/2}
\displaystyle {\sum_x}\left[ x\right]
\times \\ \times
C_{2}\left(x,c+1,d-1\right)
\left\{
\begin{array}{ccc}
k_3 &  \kappa _{12} & x \\
L_{12...c}^{\prime } &  L_{12...c} & I_1
\end{array}
\right\}
\left\{
\begin{array}{ccc}
k_3 &  \kappa _{12} & x \\
L_{12...d-1}^{\prime } &  L_{12...d-1} & I_2
\end{array}
\right\} .
\end{array}
\end{equation}
From (\ref{eq:mf}) we have that in the expressions (\ref{eq:keturi}), (\ref
{eq:r-c}), (\ref{eq:r-d}) and (\ref{eq:r-van})
the ranks $k_1=l_\alpha $, $k_2=l_\beta $, $%
k_3=l_\gamma $, $k_4=l_\delta $.

When the total rank $k=0$ and $\kappa _{12}=\kappa _{12}^{\prime }=k$, the
recoupling matrix has the form:
\begin{equation}
\label{eq:rp}
\begin{array}[b]{c}
R\left( l_a,L_a,l_b,L_b,l_c,L_c,l_d,L_d,k_1,k_2,k,k_3,k_4,k,0\right) = \\
=\left[ L_a,L_b,L_c,L_d^{\prime },k\right] ^{-1/2}\delta \left(
L_1,L_1^{\prime }\right) ...\delta \left( L_{a-1},L_{a-1}^{\prime }\right)
\times \\
\times \delta \left( L_{a+1},L_{a+1}^{\prime }\right) ...\delta \left(
L_{b-1},L_{b-1}^{\prime }\right) \delta \left( L_{b+1},L_{b+1}^{\prime
}\right) ...\times \\
\times ...\delta \left( L_{c-1},L_{c-1}^{\prime }\right) \delta \left(
L_{c+1},L_{c+1}^{\prime }\right) ...\delta \left( L_{d-1},L_{d-1}^{\prime
}\right) \times \\
\times \delta \left( L_{d+1},L_{d+1}^{\prime }\right) ...\delta \left(
L_u,L_u^{\prime }\right) \times \\
\times \delta \left( L_{12},L_{12}^{\prime }\right) ...\delta \left(
L_{12...a-1},L_{12...a-1}^{\prime }\right) \delta \left(
L_{12...d},L_{12...d}^{\prime }\right) ...\delta \left( L,L^{\prime }\right)
\times \\
\\
\times \left\{
\begin{array}{ll}
\begin{array}{l}
C_4\left( k_1,k_2,k,1\right) C_2\left( k,b+1,c-1\right) \times \\
\times C_4\left( k,k_3,k_4,1\right) C_2\left( k_4,c+1,d-1\right)
\times C_5\left( 3\right) ;
\end{array}
& \mbox{ for } a=1,b=2 \\ \ \\
\begin{array}{l}
C_1C_2\left( k,a+1,b-1\right) C_4\left( k_1,k_2,k,1\right) \times \\
\times C_2\left( k,b+1,c-1\right) C_4\left( k,k_3,k_4,2\right) \times \\
\times C_2\left( k_4,c+1,d-1\right) C_5\left( 3\right) ;
\end{array}
& \mbox{ for } a<3 \\ \ \\
\begin{array}{l}
\delta \left( L_{12},L_{12}^{\prime }\right) ...\delta \left(
L_{12...a-1},L_{12...a-1}^{\prime }\right) C_1\times \\
\times C_2\left( k,a+1,b-1\right) C_4\left( k_1,k_2,k,1\right) \times \\
\times C_2\left( k,b+1,c-1\right) C_4\left( k,k_3,k_4,2\right) \times \\
\times C_2\left( k_4,c+1,d-1\right) C_5\left( 3\right) .
\end{array}
& \mbox{ for } a\geq 3
\end{array}
\right.
\end{array}
\end{equation}

Expression (\ref{eq:rp}) have also no analogue in Cowan (1981).

Thus, we have studied all possible cases of matrix elements of arbitrary
two-electron operators. The expressions for recoupling matrices ((\ref{eq:rl}%
), (\ref{eq:ro}) and (\ref{eq:rp})) obtained in this section are simpler
and, thus, more convenient for practical applications, than those of Cowan
(1981), except for the simplest case $k=0$ of an operator acting
on one shell (\ref {eq:rg}), where they are equivalent.

\section{\bf Calculation of tensorial quantities}

In this section we will consider the submatrix elements
$$
T\left( n_i\lambda _i,n_j\lambda _j,n_i^{\prime }\lambda _i^{\prime
},n_j^{\prime }\lambda _j^{\prime },\Lambda ^{bra},\Lambda ^{ket},\Xi
,\Gamma \right)
$$
appearing in (\ref{eq:mg}).
Taking into account, that operators
$a_{m_\lambda }^{\left(
\lambda \right) }$ and $\stackrel{\sim }{a}_{m_\lambda }^{\left( \lambda
\right) }$ are components of the tensor $a_{m_qm_\lambda }^{\left( q\lambda
\right) }$, having in quasispin space the rank $q=\frac 12$ and
projections $m_q=\pm \frac 12$, i.e. $a_{\frac 12m_\lambda }^{\left(
q\lambda \right) }=a_{m_lm_s}^{\left( ls\right) }$ and $a_{-\frac
12m_\lambda }^{\left( q\lambda \right) }=\stackrel{\sim }{a}%
_{m_lm_s}^{\left( ls\right) }$
the operators $A^{\left( kk'\right) }\left( n\lambda ,\Xi
\right) $, $B^{\left( kk'\right) }(n\lambda ,\Xi )$, $C^{\left( kk'\right)
}(n\lambda ,\Xi )$, $D^{\left( ls\right) }$, $E^{\left( kk'\right) }(n\lambda
,\Xi )$ (see (\ref {eq:mc})-(\ref {eq:mf}))
in our case correspond, respectively, to the following five
expressions:

\begin{equation}
\label{eq:ta}a_{m_q}^{\left( q\lambda \right) },
\end{equation}

\begin{equation}
\label{eq:tb}\left[ a_{m_{q1}}^{\left( q\lambda \right) }\times
a_{m_{q2}}^{\left( q\lambda \right) }\right] ^{\left( \kappa _1\sigma
_1\right) },
\end{equation}

\begin{equation}
\label{eq:tc}\left[ a_{m_{q1}}^{\left( q\lambda \right) }\times \left[
a_{m_{q2}}^{\left( q\lambda \right) }\times a_{m_{q3}}^{\left( q\lambda
\right) }\right] ^{\left( \kappa _1\sigma _1\right) }\right] ^{\left( \kappa
_2\sigma _2\right) },
\end{equation}

\begin{equation}
\label{eq:td}\left[ \left[ a_{m_{q1}}^{\left( q\lambda \right) }\times
a_{m_{q2}}^{\left( q\lambda \right) }\right] ^{\left( \kappa _1\sigma
_1\right) }\times a_{m_{q3}}^{\left( q\lambda \right) }\right] ^{\left(
\kappa _2\sigma _2\right) },
\end{equation}

\begin{equation}
\label{eq:te}\left[ \left[ a_{m_{q1}}^{\left( q\lambda \right) }\times
a_{m_{q2}}^{\left( q\lambda \right) }\right] ^{\left( \kappa _1\sigma
_1\right) }\times \left[ a_{m_{q3}}^{\left( q\lambda \right) }\times
a_{m_{q4}}^{\left( q\lambda \right) }\right] ^{\left( \kappa _2\sigma
_2\right) }\right] ^{\left( kk\right) }.
\end{equation}
We will discuss the derivation of submatrix elements of
these operators, and present the expressions for
these quantities. It is worth noting that these tensorial quantities all act
upon the {\em same} shell. So, all the advantages of tensor algebra and the
quasispin formalism may be exploited efficiently.

We obtain the submatrix elements of operator (\ref{eq:ta}) by
straightforwardly using the Wigner-Eckart theorem in quasispin space:
\begin{equation}
\label{eq:tf}
\begin{array}[b]{c}
\left( l^N\;\alpha QLS||a_{m_q}^{\left( qls\right) }||l^{N^{\prime
}}\;\alpha ^{\prime }Q^{\prime }L^{\prime }S^{\prime }\right) =-\left[
Q\right] ^{-1/2}\left[
\begin{array}{ccc}
Q^{\prime } & 1/2 & Q \\
M_Q^{\prime } & m_q & M_Q
\end{array}
\right] \times \\
\times \left( l\;\alpha QLS|||a^{\left( qls\right) }|||l\;\alpha ^{\prime
}Q^{\prime }L^{\prime }S^{\prime }\right) ,
\end{array}
\end{equation}
where the last multiplier in (\ref{eq:tf}) is the so-called completely
reduced (reduced in the quasispin, orbital and spin spaces) matrix element.
The coefficient
$$\left[ \begin{array}{ccc}
j_1 & j_2 & j \\
m_1 & m_2 & m
\end{array}
\right] $$ is a Clebsch-Gordan coefficient. Different notations for it occur,
e.g. $A^{j j_1 j_2} _{m m_1 m_2}$ in Eckart (1930), $S^{j_1 j_2} _{j m_1 m_2}$
in Wigner (1931), $(j_1 j_2 m_1 m_2 | j_1 j_2 j m)$ in Condon and Shortley
(1935) or Judd (1967).

The value of the submatrix element of operator (\ref{eq:tb}) is obtained by
basing our development on (33), (34) of Gaigalas and Rudzikas (1996). In the
other three cases (\ref{eq:tc}), (\ref{eq:td}), (\ref{eq:te}) we obtain them
by using (2.28) of Jucys and Savukynas (1973):
\begin{equation}
\label{eq:tg}
\begin{array}[b]{c}
(nl^N\;\alpha QLS||\left[ F^{\left( \kappa _1\sigma _1\right) }\left(
n\lambda \right) \times G^{(\kappa _2\sigma _2)}\left( n\lambda \right)
\right] ^{\left( kk\right) }||nl^{N^{\prime }}\;\alpha ^{\prime }Q^{\prime
}L^{\prime }S^{\prime })= \\
=\left( -1\right) ^{L+S+L^{\prime }+S^{\prime }+2k}\left[ k\right] \times \\
\times
\displaystyle {\sum_{\alpha ^{\prime \prime }Q^{\prime \prime }L^{\prime
\prime }S^{\prime \prime }}}(nl^N\;\alpha QLS||F^{\left( \kappa _1\sigma
_1\right) }\left( n\lambda \right) ||nl^{N^{\prime \prime }}\;\alpha
^{\prime \prime }Q^{\prime \prime }L^{\prime \prime }S^{\prime \prime
})\times \\ \times (nl^{N^{\prime \prime }}\;\alpha ^{\prime \prime
}Q^{\prime \prime }L^{\prime \prime }S^{\prime \prime }||G^{(\kappa _2\sigma
_2)}\left( n\lambda \right) ||nl^{N^{\prime }}\;\alpha ^{\prime }Q^{\prime
}L^{\prime }S^{\prime })\times \\
\times \left\{
\begin{array}{ccc}
\kappa _1 & \kappa _2 & k \\
L^{\prime } & L & L^{\prime \prime }
\end{array}
\right\} \left\{
\begin{array}{ccc}
\sigma _1 & \sigma _2 & k \\
S^{\prime } & S & S^{\prime \prime }
\end{array}
\right\} ,
\end{array}
\end{equation}
where $F^{\left( \kappa _1\sigma _1\right) }\left( n\lambda \right) $, $%
G^{(\kappa _2\sigma _2)}\left( n\lambda \right) $ is one of (\ref{eq:ta}) or
(\ref{eq:tb}) and the submatrix elements correspondingly are defined by (\ref
{eq:tf}) and (33), (34) of Gaigalas and Rudzikas (1996). $N^{\prime \prime }$
is defined by the second quantization operators occurring in $F^{\left(
\kappa _1\sigma _1\right) }\left( n\lambda \right) $ and $G^{(\kappa
_2\sigma _2)}\left( n\lambda \right) $.

As is seen, by using this approach, the calculation of the angular parts of
matrix elements between functions with $u$ open shells is
reduced to requiring the
submatrix elements of tensors (\ref{eq:ta}) and (\ref{eq:tb}) within one shell
of equivalent electrons. As these completely reduced submatrix elements do
not depend on the occupation number of the shell, the tables for
these quantities are considerably reduced in size in comparison with
the tables of analogous submatrix
elements of tensorial quantities $U^k,$ $V^{k_1k_2}$ (Jucys and Savukynas
1973) and the tables of fractional parentage coefficients.

\section{\bf Phase factor}

In this section we present the phase factors $\Delta $ in (\ref{eq:mg}),
which appear for submatrix elements of operators
in Eq. (\ref{eq:mc})-(\ref{eq:mf}).

For distributions 1-6 (Table \ref{pasis}):
\begin{equation}
\label{eq:ti}\Delta =0.
\end{equation}
For the distributions 7-18 (Table \ref{pasis}):
\begin{equation}
\label{eq:tj}\Delta =1+\stackrel{j-1}{\sum_{k=i}}N_k,
\end{equation}
where if $\alpha <\beta $, then $i=\alpha ,$ $j=\beta $, and if $\alpha
>\beta $, then $i=\beta ,$ $j=\alpha $; $N_k$ is the occupation number of a
shell of equivalent electrons having the label $k$. For the distributions
19-42 (Table \ref{pasis}):
\begin{equation}
\label{eq:tl}\Delta =\stackrel{\beta -1}{\sum_{k=\alpha }}N_k+\stackrel{%
\delta -1}{\sum_{k=\gamma }}N_k.
\end{equation}

\section{\bf Spin-angular part of any two-particle operator}

{%
\begin{table}
\begin{center}
\caption{Scheme of the expressions for matrix elements of any two-particle
operator}
\label{israis}
  \begin{tabular}{|r|c|c|c|c|c|c|c|c|c|} \hline
 & & & & & & & & & \\
    {\bf No.} & {\bf $\widehat{G}$ } & {\bf $\widehat{G}(T)$ } &
     {\bf $\alpha$ } & {\bf $\beta$ }
& {\bf $\gamma$ } & {\bf $\delta$ }& $ \tilde \Theta$ &
{\bf $R$ }& $\Delta$ \\
 & & & & & & & & & \\ \hline \hline
{\bf 1.}&(\ref{eq:ad-aa}) &(\ref{eq:mc})&(\ref{eq:te})& --& --& --
&(\ref{eq:ad-ab})&(\ref{eq:rc}), (\ref{eq:rg})&(\ref{eq:ti})\\
   &(\ref{eq:ad-ad}) &(\ref{eq:mc})&(\ref{eq:te})& --& --& --
&(\ref{eq:ad-ae}), (\ref{eq:ad-af})&(\ref{eq:rc}), (\ref{eq:rg})&(\ref{eq:ti})\\
 \hline \hline
{\bf 2.}&(\ref{eq:ab-a})  &(\ref{eq:md})&(\ref{eq:tb})&(\ref{eq:tb})& --& --
&(\ref{eq:ab})    &(\ref{eq:rh}), (\ref{eq:rl})&(\ref{eq:ti})\\
{\bf 3.}&(\ref{eq:ab-a})  &(\ref{eq:md})&(\ref{eq:tb})&(\ref{eq:tb})& --& --
&(\ref{eq:ab})    &(\ref{eq:rh}), (\ref{eq:rl})&(\ref{eq:ti})\\
{\bf 4.}&(\ref{eq:ad-a})  &(\ref{eq:md})&(\ref{eq:tb})&(\ref{eq:tb})& --& --
&(\ref{eq:ad})    &(\ref{eq:rh}), (\ref{eq:rl})&(\ref{eq:ti})\\
{\bf 5.}&(\ref{eq:ad-a})  &(\ref{eq:md})&(\ref{eq:tb})&(\ref{eq:tb})& --& --
&(\ref{eq:ad})    &(\ref{eq:rh}), (\ref{eq:rl})&(\ref{eq:ti})\\
 \hline
{\bf 6.}&(\ref{eq:ac-a})  &(\ref{eq:md})&(\ref{eq:tb})&(\ref{eq:tb})& --& --
&(\ref{eq:ac})    &(\ref{eq:rh}), (\ref{eq:rl})&(\ref{eq:ti})\\
 \hline
{\bf 7.}&(\ref{eq:ae-a})  &(\ref{eq:md})&(\ref{eq:tc})&(\ref{eq:ta})& --& --
&(\ref{eq:ae-b})  &(\ref{eq:rh}), (\ref{eq:rl})&(\ref{eq:tj})\\
{\bf 8.}&(\ref{eq:ae-a})  &(\ref{eq:md})&(\ref{eq:tc})&(\ref{eq:ta})& --& --
&(\ref{eq:ae})    &(\ref{eq:rh}), (\ref{eq:rl})&(\ref{eq:tj})\\
{\bf 9.}&(\ref{eq:af-a})  &(\ref{eq:md})&(\ref{eq:ta})&(\ref{eq:td})& --& --
&(\ref{eq:af-b})  &(\ref{eq:rh}), (\ref{eq:rl})&(\ref{eq:tj})\\
{\bf 10.}&(\ref{eq:af-a})  &(\ref{eq:md})&(\ref{eq:ta})&(\ref{eq:td})& --& --
&(\ref{eq:af})    &(\ref{eq:rh}), (\ref{eq:rl})&(\ref{eq:tj})\\
 \hline \hline
{\bf 11.}&(\ref{eq:ab-a})  &(\ref{eq:me})&(\ref{eq:ta})&(\ref{eq:ta})&(\ref{eq:tb})& --
&(\ref{eq:ab})    &(\ref{eq:rn}), (\ref{eq:ro})&(\ref{eq:tj})\\
{\bf 12.}&(\ref{eq:ab-a})  &(\ref{eq:me})&(\ref{eq:ta})&(\ref{eq:ta})&(\ref{eq:tb})& --
&(\ref{eq:ab})    &(\ref{eq:rn}), (\ref{eq:ro})& (\ref{eq:tj})\\
{\bf 13.}&(\ref{eq:ad-a})  &(\ref{eq:me})&(\ref{eq:ta})&(\ref{eq:ta})&(\ref{eq:tb})& --
&(\ref{eq:ad})    &(\ref{eq:rn}), (\ref{eq:ro})& (\ref{eq:tj})\\
{\bf 14.}&(\ref{eq:ad-a})  &(\ref{eq:me})&(\ref{eq:ta})&(\ref{eq:ta})&(\ref{eq:tb})& --
&(\ref{eq:ad})    &(\ref{eq:rn}), (\ref{eq:ro})& (\ref{eq:tj})\\
 \hline
{\bf 15.}&(\ref{eq:ac-a})&(\ref{eq:me})&(\ref{eq:ta})&(\ref{eq:ta})&(\ref{eq:tb})& --
&(\ref{eq:ac})    &(\ref{eq:rn}), (\ref{eq:ro})&(\ref{eq:tj})\\
{\bf 16.}&(\ref{eq:ac-a})&(\ref{eq:me})&(\ref{eq:ta})&(\ref{eq:ta})&(\ref{eq:tb})& --
&(\ref{eq:ac})    &(\ref{eq:rn}), (\ref{eq:ro})&(\ref{eq:tj})\\
 \hline
{\bf 17.}&(\ref{eq:ac-a})&(\ref{eq:me})&(\ref{eq:ta})&(\ref{eq:ta})&(\ref{eq:tb})& --
&(\ref{eq:ac})    &(\ref{eq:rn}), (\ref{eq:ro})&(\ref{eq:tj})\\
{\bf 18.}&(\ref{eq:ac-a})&(\ref{eq:me})&(\ref{eq:ta})&(\ref{eq:ta})&(\ref{eq:tb})& --
&(\ref{eq:ac})    &(\ref{eq:rn}), (\ref{eq:ro})&(\ref{eq:tj})\\
 \hline \hline
{\bf 19.}&(\ref{eq:ac-a})&(\ref{eq:mf})&(\ref{eq:ta})&(\ref{eq:ta})&(\ref{eq:ta})&(\ref{eq:ta})
&(\ref{eq:ac})    &(\ref{eq:keturi}), (\ref{eq:rp})&(\ref{eq:tl})\\
{\bf 20.}&(\ref{eq:ac-a})&(\ref{eq:mf})&(\ref{eq:ta})&(\ref{eq:ta})&(\ref{eq:ta})&(\ref{eq:ta})
&(\ref{eq:ac})    &(\ref{eq:keturi}), (\ref{eq:rp})&(\ref{eq:tl})\\
{\bf 21.}&(\ref{eq:ac-a})&(\ref{eq:mf})&(\ref{eq:ta})&(\ref{eq:ta})&(\ref{eq:ta})&(\ref{eq:ta})
&(\ref{eq:ac})    &(\ref{eq:keturi}), (\ref{eq:rp})&(\ref{eq:tl})\\
{\bf 22.}&(\ref{eq:ac-a})&(\ref{eq:mf})&(\ref{eq:ta})&(\ref{eq:ta})&(\ref{eq:ta})&(\ref{eq:ta})
&(\ref{eq:ac})    &(\ref{eq:keturi}), (\ref{eq:rp})&(\ref{eq:tl}) \\
\hline
\end{tabular}
\end{center}
\end{table}
}

\clearpage
\vspace {0.5in} Table 8 (continued) \vspace {0.1in}

\begin{center}
\begin{tabular}{|r|c|c|c|c|c|c|c|c|c|} \hline
 & & & & & & & & & \\
    {\bf No.} & {\bf $\widehat{G}$ } & {\bf $\widehat{G}(T)$ } &
    {\bf $\alpha$ } & {\bf $\beta$ }
& {\bf $\gamma$ } & {\bf $\delta$ }& $ \tilde \Theta$ & {\bf $R$ }& $\Delta$ \\
 & & & & & & & & & \\  \hline \hline
{\bf 23.}&(\ref{eq:ac-a})&(\ref{eq:mf})&(\ref{eq:ta})&(\ref{eq:ta})&(\ref{eq:ta})&(\ref{eq:ta})
&(\ref{eq:ac})    &(\ref{eq:keturi}), (\ref{eq:rp})&(\ref{eq:tl})\\
{\bf 24.}&(\ref{eq:ac-a})&(\ref{eq:mf})&(\ref{eq:ta})&(\ref{eq:ta})&(\ref{eq:ta})&(\ref{eq:ta})
&(\ref{eq:ac})    &(\ref{eq:keturi}), (\ref{eq:rp})&(\ref{eq:tl})\\
{\bf 25.}&(\ref{eq:ac-a})&(\ref{eq:mf})&(\ref{eq:ta})&(\ref{eq:ta})&(\ref{eq:ta})&(\ref{eq:ta})
&(\ref{eq:ac})    &(\ref{eq:keturi}), (\ref{eq:rp})&(\ref{eq:tl})\\
{\bf 26.}&(\ref{eq:ac-a})&(\ref{eq:mf})&(\ref{eq:ta})&(\ref{eq:ta})&(\ref{eq:ta})&(\ref{eq:ta})
&(\ref{eq:ac})    &(\ref{eq:keturi}), (\ref{eq:rp})&(\ref{eq:tl}) \\
 \hline
{\bf 27.}&(\ref{eq:ab-a})&(\ref{eq:mf})&(\ref{eq:ta})&(\ref{eq:ta})&(\ref{eq:ta})&(\ref{eq:ta})
&(\ref{eq:ab})    &(\ref{eq:keturi}), (\ref{eq:rp})&(\ref{eq:tl})\\
{\bf 28.}&(\ref{eq:ad-a})&(\ref{eq:mf})&(\ref{eq:ta})&(\ref{eq:ta})&(\ref{eq:ta})&(\ref{eq:ta})
&(\ref{eq:ad})    &(\ref{eq:keturi}), (\ref{eq:rp})&(\ref{eq:tl})\\
{\bf 29.}&(\ref{eq:ab-a})&(\ref{eq:mf})&(\ref{eq:ta})&(\ref{eq:ta})&(\ref{eq:ta})&(\ref{eq:ta})
&(\ref{eq:ab})    &(\ref{eq:keturi}), (\ref{eq:rp})&(\ref{eq:tl})\\
{\bf 30.}&(\ref{eq:ad-a})&(\ref{eq:mf})&(\ref{eq:ta})&(\ref{eq:ta})&(\ref{eq:ta})&(\ref{eq:ta})
&(\ref{eq:ad})    &(\ref{eq:keturi}), (\ref{eq:rp})&(\ref{eq:tl})\\
 \hline
{\bf 31.}&(\ref{eq:ab-a})&(\ref{eq:mf})&(\ref{eq:ta})&(\ref{eq:ta})&(\ref{eq:ta})&(\ref{eq:ta})
&(\ref{eq:ab})    &(\ref{eq:keturi}), (\ref{eq:rp})&(\ref{eq:tl})\\
{\bf 32.}&(\ref{eq:ab-a})&(\ref{eq:mf})&(\ref{eq:ta})&(\ref{eq:ta})&(\ref{eq:ta})&(\ref{eq:ta})
&(\ref{eq:ab})    &(\ref{eq:keturi}), (\ref{eq:rp})&(\ref{eq:tl})\\
{\bf 33.}&(\ref{eq:ad-a})&(\ref{eq:mf})&(\ref{eq:ta})&(\ref{eq:ta})&(\ref{eq:ta})&(\ref{eq:ta})
&(\ref{eq:ad})    &(\ref{eq:keturi}), (\ref{eq:rp})&(\ref{eq:tl})\\
{\bf 34.}&(\ref{eq:ad-a})&(\ref{eq:mf})&(\ref{eq:ta})&(\ref{eq:ta})&(\ref{eq:ta})&(\ref{eq:ta})
&(\ref{eq:ad})    &(\ref{eq:keturi}), (\ref{eq:rp})&(\ref{eq:tl})\\
 \hline
{\bf 35.}&(\ref{eq:ab-a})&(\ref{eq:mf})&(\ref{eq:ta})&(\ref{eq:ta})&(\ref{eq:ta})&(\ref{eq:ta})
&(\ref{eq:ab})    &(\ref{eq:keturi}), (\ref{eq:rp})&(\ref{eq:tl})\\
{\bf 36.}&(\ref{eq:ab-a})&(\ref{eq:mf})&(\ref{eq:ta})&(\ref{eq:ta})&(\ref{eq:ta})&(\ref{eq:ta})
&(\ref{eq:ab})    &(\ref{eq:keturi}), (\ref{eq:rp})&(\ref{eq:tl})\\
{\bf 37.}&(\ref{eq:ad-a})&(\ref{eq:mf})&(\ref{eq:ta})&(\ref{eq:ta})&(\ref{eq:ta})&(\ref{eq:ta})
&(\ref{eq:ad})    &(\ref{eq:keturi}), (\ref{eq:rp})&(\ref{eq:tl})\\
{\bf 38.}&(\ref{eq:ad-a})&(\ref{eq:mf})&(\ref{eq:ta})&(\ref{eq:ta})&(\ref{eq:ta})&(\ref{eq:ta})
&(\ref{eq:ad})    &(\ref{eq:keturi}), (\ref{eq:rp})&(\ref{eq:tl})\\
 \hline
{\bf 39.}&(\ref{eq:ab-a})&(\ref{eq:mf})&(\ref{eq:ta})&(\ref{eq:ta})&(\ref{eq:ta})&(\ref{eq:ta})
&(\ref{eq:ab})    &(\ref{eq:keturi}), (\ref{eq:rp})&(\ref{eq:tl})\\
{\bf 40.}&(\ref{eq:ab-a})&(\ref{eq:mf})&(\ref{eq:ta})&(\ref{eq:ta})&(\ref{eq:ta})&(\ref{eq:ta})
&(\ref{eq:ab})    &(\ref{eq:keturi}), (\ref{eq:rp})&(\ref{eq:tl})\\
{\bf 41.}&(\ref{eq:ad-a})&(\ref{eq:mf})&(\ref{eq:ta})&(\ref{eq:ta})&(\ref{eq:ta})&(\ref{eq:ta})
&(\ref{eq:ad})    &(\ref{eq:keturi}), (\ref{eq:rp})&(\ref{eq:tl})\\
{\bf 42.}&(\ref{eq:ad-a})&(\ref{eq:mf})&(\ref{eq:ta})&(\ref{eq:ta})&(\ref{eq:ta})&(\ref{eq:ta})
&(\ref{eq:ad})    &(\ref{eq:keturi}), (\ref{eq:rp})&(\ref{eq:tl})\\     \hline
\end{tabular}
\end{center}

%\vspace {0.5in}

In the previous sections, all the expressions required to calculate
the spin-angular part of any two-particle operator were given.
For convenience, the structure of the expressions (the numbers of the
corresponding formulas) are summarized in the Table \ref{israis} for each
distribution given in the Table \ref{pasis}. The classification numbers of
the distributions are presented in the first column of the Table \ref{israis}.
The equation number of the tensorial expression of the
two-particle operator $\widehat{G}$  is given in the second column,
and  the equation number of the
tensorial class of the two-particle operator, denoted by
$\widehat{G}(T)$, in the third column.

In the next four columns are given the numbers of the formulas of the
tensors, which act inside the shell. A tensor acting upon the $\alpha $ shell is
given in the $\alpha $ column, and in the columns $\beta $, $\gamma $, $%
\delta $ - upon the $\beta $, $\gamma $, $\delta $ shells, respectively.
Consequently, if we want to find submatrix element,
$T\left( n_i\lambda _i,n_j\lambda _j,n_i^{\prime }\lambda _i^{\prime
},n_j^{\prime }\lambda _j^{\prime },\Lambda ^{bra},\Lambda ^{ket},\Xi
,\Gamma \right) $, at first we have to calculate the submatrix element of the
tensor from column $\alpha $ between functions, consisting only of the
$\alpha $
shell, and then to look for the submatrix element of the tensor from column $%
\beta $ between functions, consisting only of $\beta $ shell, and so on.
Thus, we need to calculate only submatrix elements of the tensors acting
upon a certain shell. The details of the calculation of these submatrix
elements were discussed in the Section 5.

The coefficients $\tilde \Theta $ are given in the $\tilde \Theta $ column$.$
The numbers of expressions for the recoupling matrix $R\left( \lambda
_i,\lambda _j,\lambda _i^{\prime },\lambda _j^{\prime },\Lambda
^{bra},\Lambda ^{ket},\Gamma \right) $ and phase factor $\Delta $ are given
in the last two columns. From this table it is easy to derive the general
formulas for spin-angular parts of matrix elements of any
two-particle operator.

\section{\bf Conclusions}

The approach to matrix element evaluation that we present, is based on the
the combination of the angular momentum theory as described in Jucys and
Bandzaitis (1977), on the
concept of irreducible tensorial sets (Judd 1967, Rudzikas and Kaniauskas
1984), on a generalized graphical approach (Gaigalas {\it et al} 1985), on
the quasispin approach (Rudzikas and Kaniauskas 1984), and on the use of
reduced coefficients of fractional parentage (Rudzikas 1991,
Rudzikas 1997, Judd 1996).
All this, in its entity, introduces  a number of new features, in
comparison with traditional approaches:

(1) The tensorial expressions of a two-particle operator, presented
in Section 2, allow one to exploit all the advantages of a new version of
Racah algebra based on quasispin formalism when the latter is applied
within each particular shell only. In particular, this
is not only
a reformulation of spin-angular  calculations  in terms of
standard quantities, but also the determination beforehand
from symmetry properties, of which
matrix elements are equal to zero without performing further
explicit calculations. That is determined from the submatrix
elements $T\left(
n_i\lambda _i,n_j\lambda _j,n_i^{\prime }\lambda _i^{\prime },n_j^{\prime
}\lambda _j^{\prime },\Lambda ^{bra},\Lambda ^{ket},\Xi ,\Gamma \right)$.

(2) It enables one to use the Wigner-Eckart theorem in quasispin space. This
provides an opportunity to use tables of reduced coefficients of
fractional parentage and tables of other standard quantities (Section 5),
which do not depend on the occupation number of a shell of equivalent
electrons. Thus, the volume of tables of standard quantities is reduced
considerably in comparison with the analogous tables of submatrix elements
of tensorial operators $U^k$, $V^{k1}$ and the tables of fractional
parentage coefficients. This undoubtedly makes the inclusion of shells
of equivalent $f$ electrons with arbitrary occupation numbers
considerably easier, and the process of selecting the standard
quantities from the tables becomes simpler.

(3) The tensorial form of any operator presented in Section 2 allows one to
obtain simple expressions for the recoupling matrices (Section 4). Hence,
the computer code based on this approach would use immediately the
analytical formulas for recoupling matrices
$R\left( \lambda _i,\lambda _j,\lambda _i^{\prime },\lambda _j^{\prime
},\Lambda ^{bra},\Lambda ^{ket},\Gamma \right)$.
This feature also saves computing
time, because i) complex calculations leading finally to simple analytical
expressions (Bar-Shalom and Klapisch 1988) are avoided, and ii) a number of
momenta triads (triangular conditions) can be checked before the explicit
calculation of a recoupling matrix leading to a zero value. These triangular
conditions may be determined not only for the terms of shells that the
operators of second quantization act upon, as is the case
for the submatrix elements
$T\left(
n_i\lambda _i,n_j\lambda _j,n_i^{\prime }\lambda _i^{\prime },n_j^{\prime
}\lambda _j^{\prime },\Lambda ^{bra},\Lambda ^{ket},\Xi ,\Gamma \right)$
(see conclusion 1), but also for the rest of the shells and resulting terms.

In this approach both diagonal and non-diagonal with respect to
configurations, matrix elements are
considered in a uniform way, and are expressed in terms of the same
quantities. The difference is only in the values of the projections
of the quasispin momenta of separate shells.

In this paper all the expressions needed in the spin-angular parts of
matrix elements of two-particle operators calculation are presented. This approach is
applicable to one-particle operators as well. While calculating the
spin-angular parts of the latter, all the expressions needed are included in
the cases discussed for the two-particle operator. For instance, in the
recoupling matrix calculation two of the four cases discussed above appear,
namely, when all the second quantization operators act upon the same
shell (Section 4.1) and
when they act upon two different shells of equivalent electrons (Section
4.2). Thus, this approach is applicable to any one- and two-particle
operator. Practical usage shows that a series of
difficulties persisting in the traditional approach to the calculation of
angular parts of matrix elements based on the usage of
coefficients of fractional parentage and unit tensors can be avoided
and high efficiency may be achieved. Indeed, preliminary
calculations show that computer programs based on our approach on
average are 4-6 times faster than the other well-known codes (Gaigalas
{\it et al }1995). This methodology can easily be generalized
to cover the case of relativistic operators and relativistic
wave functions.

\section*{\bf Acknowledgements}

This work is part of a co-operative research project funded by National
Science Foundation under grant No. PHY-9501830 and by EURONET PECAM
associated contract ERBCIPDCT 940025. One of the authors (CFF) was supported
by a grant from the Division of Chemical Sciences, Office of Basic Energy
Sciences, Office of Energy Research, U.S. Department of Energy.

%\clearpage

%\clearpage

%\clearpage

\section*{Appendix}

Here algebraic expressions are presented for the two-particle operator
(\ref{eq:pirga}) in the irreducible
tensorial form for
all the distributions from Table \ref
{israis}. Although there are quite a few distributions, the
structure of their algebraic formulas is similar, and therefore on the basis
of a graphical approach (Gaigalas {\it et al} 1985)
the expressions may be written down in a compact form,
where one general formula includes all the cases having the same structure.
Each particular formula is obtained from these by performing elementary
graphical transformations according to the rules explained below. The
general expressions are:

\begin{figure}
\setlength{\unitlength}{1mm}
\begin{picture}(148,160)
\thicklines
%
% Diagram A^{1}
%
\put(10,140){\vector(0,1){10}}
\put(10,150){\vector(0,1){2}}
\put(5,148){\makebox(0,0)[t]{$ n_{\alpha}\lambda_{\alpha}$}}
\put(20,140){\vector(0,1){10}}
\put(20,150){\vector(0,1){2}}
\put(25,148){\makebox(0,0)[t]{$ n_{\alpha}\lambda_{\alpha}$}}
\multiput(11,140)(2,0){5}{\oval(2,1)[t]}
\put(15,143){\makebox(0,0){$kk$}}
\put(10,130){\line(0,1){10}}
\put(10,130){\vector(0,1){3}}
\put(5,133){\makebox(0,0){$ n_{\alpha}\lambda_{\alpha}$}}
\put(10,128){\vector(0,1){3}}
\put(20,130){\line(0,1){10}}
\put(20,130){\vector(0,1){3}}
\put(25,133){\makebox(0,0){$ n_{\alpha}\lambda_{\alpha}$}}
\put(20,128){\vector(0,1){3}}
\put(15,115){\makebox(0,0){$ A_{1}$}}
%
% Diagram A^{2}
%
\put(35,140){\makebox(0,0){$=$}}
\put(50,140){\vector(0,1){10}}
\put(50,150){\vector(0,1){2}}
\put(45,148){\makebox(0,0)[t]{$ n_{\alpha}\lambda_{\alpha}$}}
\put(60,140){\vector(0,1){10}}
\put(60,150){\vector(0,1){2}}
\put(65,148){\makebox(0,0)[t]{$ n_{\alpha}\lambda_{\alpha}$}}
\multiput(50,140)(1,0){10}{\circle*{0.02}}
\put(55,143){\makebox(0,0){$kk$}}
\put(50,130){\line(0,1){10}}
\put(50,130){\vector(0,1){3}}
\put(45,133){\makebox(0,0){$ n_{\alpha}\lambda_{\alpha}$}}
\put(50,128){\vector(0,1){3}}
\put(60,130){\line(0,1){10}}
\put(60,130){\vector(0,1){3}}
\put(65,133){\makebox(0,0){$ n_{\alpha}\lambda_{\alpha}$}}
\put(60,128){\vector(0,1){3}}
\put(55,115){\makebox(0,0){$ A_{2}$}}
%
% Diagram A^{3}
%
\put(75,140){\makebox(0,0){$+$}}
\put(90,140){\vector(0,1){10}}
\put(90,150){\vector(0,1){2}}
\put(85,148){\makebox(0,0)[t]{$ n_{\alpha}\lambda_{\alpha}$}}
\put(96,150){\makebox(0,0)[t]{$ n_{\alpha}\lambda_{\alpha}$}}
\multiput(90,140)(1,0){10}{\circle*{0.02}}
\put(95,137){\makebox(0,0){$kk$}}
\put(96,145){\vector(-1,0){2}}
\put(100,130){\line(0,1){10}}
\put(95,140){\oval(10,10)[t]}
\put(100,130){\vector(0,1){3}}
\put(105,133){\makebox(0,0){$ n_{\alpha}\lambda_{\alpha}$}}
\put(100,128){\vector(0,1){3}}
\put(95,115){\makebox(0,0){$ A_{3}$}}
%
% Diagram A^{4}
%
\put(10,90){\vector(0,1){10}}
\put(10,100){\vector(0,1){2}}
\put(5,98){\makebox(0,0)[t]{$ n_{i}\lambda_{i}$}}
\put(20,90){\vector(0,1){10}}
\put(20,100){\vector(0,1){2}}
\put(25,98){\makebox(0,0)[t]{$ n_{j}\lambda_{j}$}}
\multiput(10,90)(1,0){10}{\circle*{0.02}}
\put(15,93){\makebox(0,0){$kk$}}
\put(10,80){\line(0,1){10}}
\put(10,80){\vector(0,1){3}}
\put(5,83){\makebox(0,0){$ n'_{i}\lambda'_{i}$}}
\put(10,78){\vector(0,1){3}}
\put(20,80){\line(0,1){10}}
\put(20,80){\vector(0,1){3}}
\put(25,83){\makebox(0,0){$ n'_{j}\lambda'_{j}$}}
\put(20,78){\vector(0,1){3}}
\put(15,65){\makebox(0,0){$ A_{4}$}}
%
% Diagram A^{5}
%
\put(45,90){\vector(0,1){10}}
\put(45,100){\vector(0,1){2}}
\put(40,98){\makebox(0,0)[t]{$ n_{i}\lambda_{i}$}}
\put(55,90){\vector(0,1){10}}
\put(55,100){\vector(0,1){2}}
\put(60,98){\makebox(0,0)[t]{$ n_{j}\lambda_{j}$}}
\multiput(46,90)(2,0){5}{\oval(2,1)[t]}
\put(50,93){\makebox(0,0){$kk$}}
\put(45,80){\line(0,1){10}}
\put(45,80){\vector(0,1){3}}
\put(40,83){\makebox(0,0){$ n'_{i}\lambda'_{i}$}}
\put(45,78){\vector(0,1){3}}
\put(55,80){\line(0,1){10}}
\put(55,80){\vector(0,1){3}}
\put(60,83){\makebox(0,0){$ n'_{j}\lambda'_{j}$}}
\put(55,78){\vector(0,1){3}}
\put(50,65){\makebox(0,0){$ A_{5}$}}
%
% Diagram A^{6}
%
% linija i
\put(80,90){\vector(0,1){12}}
\put(80,101){\oval(5,5)[b]}
\put(80,90){\circle*{1,7}}
\put(75,98){\makebox(0,0)[t]{$ n_{i}\lambda_{i}$}}
%linja j
\put(110,90){\vector(0,1){12}}
\put(110,101){\oval(5,5)[b]}
\put(110,90){\circle*{1,7}}
\put(115,98){\makebox(0,0)[t]{$ n_{j}\lambda_{j}$}}
% forizontali linija desine
\put(80,90.6){\line(6,0){6}}
\put(80,90.3){\line(6,0){6}}
\put(80,90){\line(20,0){30}}
\put(80,89.7){\line(6,0){6}}
\put(80,89.4){\line(6,0){6}}
\put(87,93){\makebox(0,0){$\kappa_{1} \sigma_{1}$}}
% forizontali linija kaire
\put(104,90.6){\line(6,0){6}}
\put(104,90.3){\line(6,0){6}}
\put(104,89.7){\line(6,0){6}}
\put(104,89.4){\line(6,0){6}}
\put(103,93){\makebox(0,0){$\kappa_{2} \sigma_{2}$}}
%rezultatinis rangas
%\put(95,90){\circle*{1,7}}
\put(95,95){\makebox(0,0){${-}$}}
\put(98,87){\makebox(0,0){${2}$}}
\put(95.6,85){\line(0,1){5}}
\put(95.3,85){\line(0,1){5}}
\put(95,85){\line(0,1){5}}
\put(94.7,85){\line(0,1){5}}
\put(94.4,85){\line(0,1){5}}
\put(95,82){\makebox(0,0){$k k$}}
%linija i'
\put(80,80){\line(0,1){10}}
\put(74,84){\makebox(0,0){$ n'_{i}\lambda'_{i}$}}
\put(75,90){\makebox(0,0){${+}$}}
\put(82,87){\makebox(0,0){${1}$}}
\put(80,77){\vector(0,1){3}}
\put(80,79){\oval(5,5)[t]}
%linja j'
\put(110,80){\line(0,1){10}}
\put(116,84){\makebox(0,0){$ n'_{j}\lambda'_{j}$}}
\put(115,90){\makebox(0,0){${-}$}}
\put(108,87){\makebox(0,0){${3}$}}
\put(110,77){\vector(0,1){3}}
\put(110,79){\oval(5,5)[t]}
\put(95,65){\makebox(0,0){$ A_{6}$}}
%
% Diagram A^{7}
%
% linija i
\put(10,40){\vector(0,1){12}}
\put(10,51){\oval(5,5)[b]}
\put(10,40){\circle*{1,7}}
\put(5,48){\makebox(0,0)[t]{$ n_{i}\lambda_{i}$}}
%linja i' (j)
\put(40,40){\line(0,12){12}}
\put(40,52){\vector(0,-1){3}}
\put(40,51){\oval(5,5)[b]}
\put(40,40){\circle*{1,7}}
\put(45,48){\makebox(0,0)[t]{$ n'_{i}\lambda'_{i}$}}
% forizontali linija desine
\put(10,40.6){\line(6,0){6}}
\put(10,40.3){\line(6,0){6}}
\put(10,40){\line(20,0){30}}
\put(10,39.7){\line(6,0){6}}
\put(10,39.4){\line(6,0){6}}
\put(17,43){\makebox(0,0){$\kappa_{12} \sigma_{12}$}}
% forizontali linija kaire
\put(34,40.6){\line(6,0){6}}
\put(34,40.3){\line(6,0){6}}
\put(34,39.7){\line(6,0){6}}
\put(34,39.4){\line(6,0){6}}
\put(33,43){\makebox(0,0){$\kappa_{12}' \sigma_{12}' $}}
%rezultatinis rangas
\put(25,40){\circle*{1,7}}
\put(25,45){\makebox(0,0){${-}$}}
\put(28,37){\makebox(0,0){${2}$}}
\put(25.6,35){\line(0,1){5}}
\put(25.3,35){\line(0,1){5}}
\put(25,35){\line(0,1){5}}
\put(24.7,35){\line(0,1){5}}
\put(24.4,35){\line(0,1){5}}
\put(25,32){\makebox(0,0){$k k$}}
%linija j (i')
\put(10,30){\line(0,1){10}}
\put(4,34){\makebox(0,0){$ n_{j}\lambda_{j}$}}
\put(5,40){\makebox(0,0){${+}$}}
\put(12,37){\makebox(0,0){${1}$}}
\put(10,30){\vector(0,-1){3}}
\put(10,29){\oval(5,5)[t]}
%linja j'
\put(40,30){\line(0,1){10}}
\put(46,34){\makebox(0,0){$ n'_{j}\lambda'_{j}$}}
\put(45,40){\makebox(0,0){${-}$}}
\put(38,37){\makebox(0,0){${3}$}}
\put(40,27){\vector(0,1){3}}
\put(40,29){\oval(5,5)[t]}
\put(25,20){\makebox(0,0){$ A_{7}$}}
%
% Diagram A^{8}
%
% linija i
\put(80,40){\vector(0,1){12}}
\put(80,51){\oval(5,5)[b]}
\put(80,40){\circle*{1,7}}
\put(75,48){\makebox(0,0)[t]{$ n_{i}\lambda_{i}$}}
%linja j
\put(110,40){\vector(0,1){12}}
\put(110,51){\oval(5,5)[b]}
\put(110,40){\circle*{1,7}}
\put(115,48){\makebox(0,0)[t]{$ n_{j}\lambda_{j}$}}
% forizontali linija desine
\put(80,40.6){\line(6,0){6}}
\put(80,40.3){\line(6,0){6}}
\put(80,40){\line(20,0){30}}
\put(80,39.7){\line(6,0){6}}
\put(80,39.4){\line(6,0){6}}
\put(87,43){\makebox(0,0){$\kappa_{12} \sigma_{12}$}}
% forizontali linija kaire
\put(104,40.6){\line(6,0){6}}
\put(104,40.3){\line(6,0){6}}
\put(104,39.7){\line(6,0){6}}
\put(104,39.4){\line(6,0){6}}
\put(103,43){\makebox(0,0){$\kappa_{12}' \sigma_{12}' $}}
%rezultatinis rangas
\put(95,40){\circle*{1,7}}
\put(95,45){\makebox(0,0){${-}$}}
\put(98,37){\makebox(0,0){${2}$}}
\put(95.6,35){\line(0,1){5}}
\put(95.3,35){\line(0,1){5}}
\put(95,35){\line(0,1){5}}
\put(94.7,35){\line(0,1){5}}
\put(94.4,35){\line(0,1){5}}
\put(95,32){\makebox(0,0){$k k$}}
%linija j' (i')
\put(80,30){\line(0,1){10}}
\put(74,34){\makebox(0,0){$ n'_{j}\lambda'_{j}$}}
\put(75,40){\makebox(0,0){${+}$}}
\put(82,37){\makebox(0,0){${1}$}}
\put(80,27){\vector(0,1){3}}
\put(80,29){\oval(5,5)[t]}
%linja i' (j')
\put(110,30){\line(0,1){10}}
\put(116,34){\makebox(0,0){$ n'_{i}\lambda'_{i}$}}
\put(115,40){\makebox(0,0){${-}$}}
\put(108,37){\makebox(0,0){${3}$}}
\put(110,27){\vector(0,1){3}}
\put(110,29){\oval(5,5)[t]}
\put(95,20){\makebox(0,0){$ A_{8}$}}
\end{picture}
\caption{Diagrams for an arbitrary two-particle operator. Diagrams
$A_1$, $A_2$ and $A_3$ represent two-particle operators
when this
operator has distribution $\alpha \alpha \alpha \alpha$.
Diagrams $A_4$ and $A_5$ represent two-particle operators for
all others distributions.
Diagrams $A_1$, $A_2$, $A_3$, $A_4$ and $A_5$ are similar to
the usual Feynman-Goldstone diagrams.
Diagrams $A_6$, $A_7$ and $A_8$
represent tensorial products of second quantization operators
($A_6$ for the second group, $A_7$ for the third, and $A_8$ for the
fourth.).}
\label{aa-a}
\end{figure}
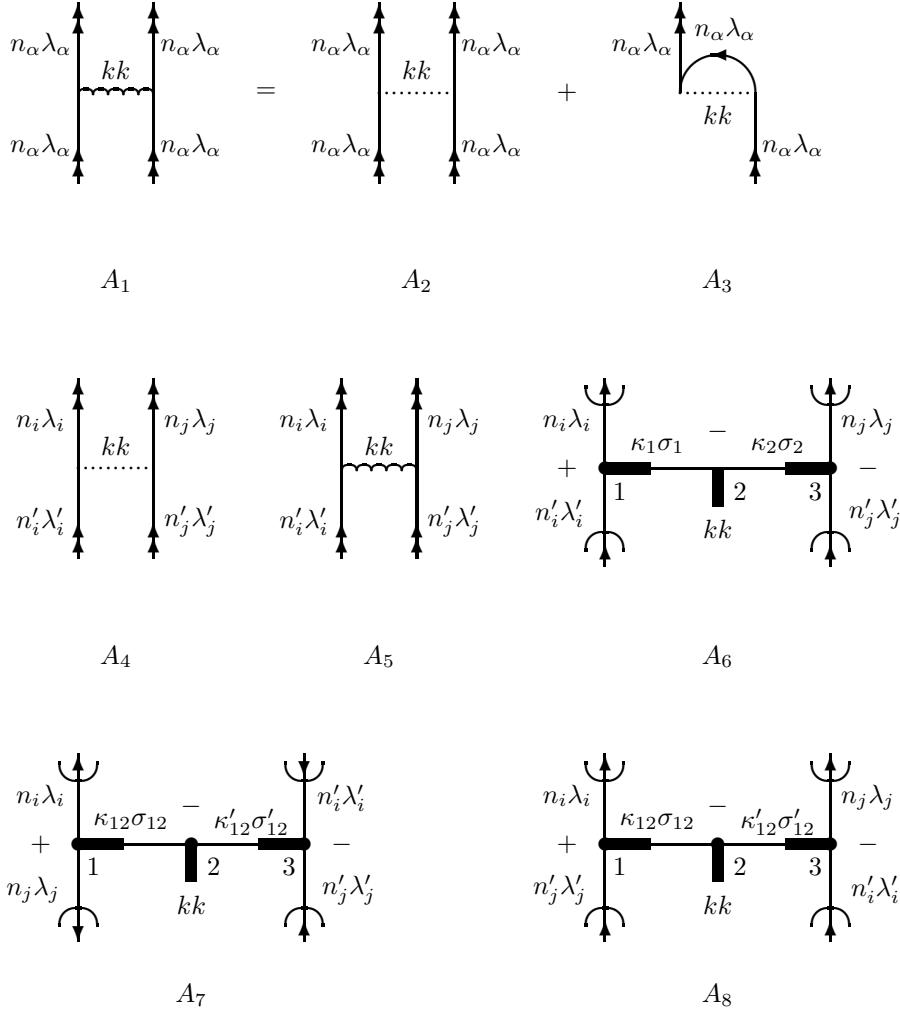

(1) Distribution $\alpha \alpha \alpha \alpha $ (case 1 from Table \ref
{israis}):

For this distribution the analytical expressions (7), (8) in Gaigalas and
Rudzikas (1996) are used, in which the quantum numbers $n_il_i,n_jl_j,n_i^{%
\prime }l_i^{\prime },n_j^{\prime }l_j^{\prime }$ acquire particular values
of a shell $\alpha $. For the first form (Figure 1, $A_1$), we
have
\begin{equation}
\label{eq:ad-aa}
\begin{array}[b]{c}
A_1=
\displaystyle {\sum_{\kappa _{12}\sigma _{12}\kappa _{12}^{\prime }\sigma
_{12}^{\prime }}}\tilde \Theta _I\left( n_\alpha \lambda _\alpha ,n_\alpha
\lambda _\alpha ,n_\alpha \lambda _\alpha ,n_\alpha \lambda _\alpha ,\Xi
\right) \times \\ \times \left[ \left[ a^{\left( l_\alpha s\right) }\times
a^{\left( l_\alpha s\right) }\right] ^{\left( \kappa _{12}\sigma
_{12}\right) }\times \left[ \tilde a^{\left( l_\alpha s\right) }\times
\tilde a^{\left( l_\alpha s\right) }\right] ^{\left( \kappa _{12}^{\prime
}\sigma _{12}^{\prime }\right) }\right] _{p,-p}^{\left( kk\right) },
\end{array}
\end{equation}
where
\begin{equation}
\label{eq:ad-ab}
\begin{array}[b]{c}
\tilde \Theta _I\left( n_\alpha \lambda _\alpha ,n_\alpha \lambda _\alpha
,n_\alpha \lambda _\alpha ,n_\alpha \lambda _\alpha ,\Xi \right) \equiv
\Theta ^{\prime }\left( n_\alpha \lambda _\alpha ,n_\alpha \lambda _\alpha
,n_\alpha \lambda _\alpha ,n_\alpha \lambda _\alpha ,\Xi \right) \equiv \\
\equiv \Theta \left( n_\alpha \lambda _\alpha ,\Xi \right)
\end{array}
\end{equation}
and
\begin{equation}
\label{eq:ad-ac}
\begin{array}[b]{c}
\tilde \Theta _I\left( n_\alpha \lambda _\alpha ,n_\alpha \lambda _\alpha
,n_\alpha \lambda _\alpha ,n_\alpha \lambda _\alpha ,\Xi \right) = \\
=\frac 12\left( -1\right) ^{k-p+1}\left[ \kappa _{12},\sigma _{12},\kappa
_{12}^{\prime },\sigma _{12}^{\prime }\right] ^{1/2}\times \\
\times \left( n_\alpha \lambda _\alpha n_\alpha \lambda _\alpha ||g^{\left(
\kappa _1\kappa _2k,\sigma _1\sigma _2k\right) }||n_\alpha \lambda _\alpha
n_\alpha \lambda _\alpha \right) \times \\
\times \left\{
\begin{array}{ccc}
l_\alpha & l_\alpha & \kappa _1 \\
l_\alpha & l_\alpha & \kappa _2 \\
\kappa _{12} & \kappa _{12}^{\prime } & k
\end{array}
\right\} \left\{
\begin{array}{ccc}
s & s & \sigma _1 \\
s & s & \sigma _2 \\
\sigma _{12} & \sigma _{12}^{\prime } & k
\end{array}
\right\}
\end{array}.
\end{equation}
In an equivalent second  form (Figure 1, $A_2+A_3$), we have
\begin{equation}
\label{eq:ad-ad}
\begin{array}[b]{c}
A_2+A_3=\tilde \Theta _{IIa}\left( n_\alpha \lambda _\alpha ,n_\alpha
\lambda _\alpha ,n_\alpha \lambda _\alpha ,n_\alpha \lambda _\alpha ,\Xi
\right) \times \\
\times \left[ \left[ a^{\left( l_\alpha s\right) }\times \tilde a^{\left(
l_\alpha s\right) }\right] ^{\left( \kappa _1\sigma _1\right) }\times \left[
\tilde a^{\left( l_\alpha s\right) }\times a^{\left( l_\alpha s\right)
}\right] ^{\left( \kappa _2\sigma _2\right) }\right] _{p,-p}^{\left(
kk\right) }+ \\
+\tilde \Theta _{IIb}\left( n_\alpha \lambda _\alpha ,n_\alpha \lambda
_\alpha ,n_\alpha \lambda _\alpha ,n_\alpha \lambda _\alpha ,\Xi \right)
\left[ a^{\left( l_\alpha s\right) }\times \tilde a^{\left( l_\alpha
s\right) }\right] _{p,-p}^{\left( kk\right) },
\end{array}
\end{equation}
where
\begin{equation}
\label{eq:ad-ae}
\begin{array}[b]{c}
\tilde \Theta _{IIa}\left( n_\alpha \lambda _\alpha ,n_\alpha \lambda
_\alpha ,n_\alpha \lambda _\alpha ,n_\alpha \lambda _\alpha ,\Xi \right)
=\frac 12\left( -1\right) ^{k-p}\left[ \kappa _1,\sigma _1,\kappa _2,\sigma
_2\right] ^{-1/2}\times \\
\times \left( n_\alpha \lambda _\alpha n_\alpha \lambda _\alpha ||g^{\left(
\kappa _1\kappa _2k,\sigma _1\sigma _2k\right) }||n_\alpha \lambda _\alpha
n_\alpha \lambda _\alpha \right)
\end{array}
\end{equation}
and
\begin{equation}
\label{eq:ad-af}
\begin{array}[b]{c}
\tilde \Theta _{IIb}\left( n_\alpha \lambda _\alpha ,n_\alpha \lambda
_\alpha ,n_\alpha \lambda _\alpha ,n_\alpha \lambda _\alpha ,\Xi \right) =
\\
=\left( -1\right) ^{k-p+1}\left( n_\alpha \lambda _\alpha n_\alpha \lambda
_\alpha ||g^{\left( \kappa _1\kappa _2k,\sigma _1\sigma _2k\right)
}||n_\alpha \lambda _\alpha n_\alpha \lambda _\alpha \right) \times \\
\times \left\{
\begin{array}{ccc}
\kappa _1 & \kappa _2 & k \\
l_\alpha & l_\alpha & l_\alpha
\end{array}
\right\} \left\{
\begin{array}{ccc}
\sigma _1 & \sigma _2 & k \\
s & s & s
\end{array}
\right\} .
\end{array}
\end{equation}
Both factors $\tilde \Theta _{IIa}\left( n_\alpha \lambda _\alpha ,n_\alpha
\lambda _\alpha ,n_\alpha \lambda _\alpha ,n_\alpha \lambda _\alpha ,\Xi
\right) $ and
$\tilde \Theta _{IIb}\left( n_\alpha \lambda _\alpha ,n_\alpha \lambda
_\alpha ,n_\alpha \lambda _\alpha ,n_\alpha \lambda _\alpha ,\Xi \right) $
have properties analogous to those of $\tilde \Theta _I$ as stated in (\ref
{eq:ad-ab}).

(2) Distributions $\alpha \beta \alpha \beta $, $\beta \alpha \beta \alpha $,
$\beta \gamma \alpha \gamma $, $\gamma \beta \gamma \alpha $, $\alpha \gamma
\beta \delta $, $\gamma \alpha \delta \beta $, $\beta \delta \alpha \gamma $%
, $\delta \beta \gamma \alpha $, $\alpha \delta \beta \gamma $, $\delta
\alpha \gamma \beta $, $\beta \gamma \alpha \delta $, $\gamma \beta \delta
\alpha $ (cases 2, 3, 11, 12, 27, 29, 31, 32, 35, 36, 39, 40 from Table \ref
{israis})
(Figure 1, $A_4$, $A_6$):
\begin{equation}
\label{eq:ab-a}A_4=\tilde \Theta \left( n_i\lambda _i,n_j\lambda
_j,n_i^{\prime }\lambda _i^{\prime },n_j^{\prime }\lambda _j^{\prime },\Xi
\right) A_6,
\end{equation}
where
\begin{equation}
\label{eq:ab}
\begin{array}[b]{c}
\tilde \Theta \left( n_i\lambda _i,n_j\lambda _j,n_i^{\prime }\lambda
_i^{\prime },n_j^{\prime }\lambda _j^{\prime },\Xi \right) =\frac 12\left(
-1\right) ^{k-p}\left[ \kappa _1,\sigma _1,\kappa _2,\sigma _2\right]
^{-1/2}\times \\
\times \left( n_i\lambda _in_j\lambda _j||g^{\left( \kappa _1\kappa
_2k,\sigma _1\sigma _2k\right) }||n_i^{\prime }\lambda _i^{\prime
}n_j^{\prime }\lambda _j^{\prime }\right) .
\end{array}
\end{equation}
Diagram $A_6$ corresponds to tensorial products of the
operators of second quantization for two-particle operator (for
details see in Gaigalas and Rudzikas 1996).

(3) Distributions $\alpha \alpha \beta \beta $, $\gamma \gamma \alpha \beta $%
, $\gamma \gamma \beta \alpha $, $\alpha \beta \gamma \gamma $, $\beta
\alpha \gamma \gamma $, $\alpha \beta \gamma \delta $, $\beta \alpha \delta
\gamma $, $\alpha \beta \delta \gamma $, $\beta \alpha \gamma \delta $, $%
\gamma \delta \alpha \beta $, $\delta \gamma \beta \alpha $, $\gamma \delta
\beta \alpha $, $\delta \gamma \alpha \beta $ (cases 6, 15-26 from Table \ref
{israis})
(Figure 1, $A_5$, $A_7$):
\begin{equation}
\label{eq:ac-a}A_5=\displaystyle {\sum_{\kappa _{12}\sigma _{12}\kappa
_{12}^{\prime }\sigma _{12}^{\prime }}}\tilde \Theta \left( n_i\lambda
_i,n_j\lambda _j,n_i^{\prime }\lambda _i^{\prime },n_j^{\prime }\lambda
_j^{\prime },\Xi \right) A_7,
\end{equation}
where
\begin{equation}
\label{eq:ac}
\begin{array}[b]{c}
\tilde \Theta \left( n_i\lambda _i,n_j\lambda _j,n_i^{\prime }\lambda
_i^{\prime },n_j^{\prime }\lambda _j^{\prime },\Xi \right) =\frac 12\left(
-1\right) ^{k-p+1}\times \\
\times \left[ \kappa _{12},\sigma _{12},\kappa _{12}^{\prime },\sigma
_{12}^{\prime }\right] ^{1/2}\left( n_i\lambda _in_j\lambda _j||g^{\left(
\kappa _1\kappa _2k,\sigma _1\sigma _2k\right) }||n_i^{\prime }\lambda
_i^{\prime }n_j^{\prime }\lambda _j^{\prime }\right) \times \\
\times \left\{
\begin{array}{ccc}
l_i & l_{^i}^{\prime } & \kappa _1 \\
l_j & l_j^{\prime } & \kappa _2 \\
\kappa _{12} & \kappa _{12}^{\prime } & k
\end{array}
\right\} \left\{
\begin{array}{ccc}
s & s & \sigma _1 \\
s & s & \sigma _2 \\
\sigma _{12} & \sigma _{12}^{\prime } & k
\end{array}
\right\} .
\end{array}
\end{equation}

(4) Distributions $\alpha \beta \beta \alpha $, $\beta \alpha \alpha \beta $,
$\gamma \beta \alpha \gamma $, $\beta \gamma \gamma \alpha $, $\alpha \gamma
\delta \beta $, $\gamma \alpha \beta \delta $, $\beta \delta \gamma \alpha $%
, $\delta \beta \alpha \gamma $, $\alpha \delta \gamma \beta $, $\delta
\alpha \beta \gamma $, $\beta \gamma \delta \alpha $, $\gamma \beta \alpha
\delta $ (cases 4, 5, 13, 14, 28, 30, 33, 34, 37, 38, 41, 42 from Table \ref
{israis})
(Figure 1, $A_4$, $A_8$):
\begin{equation}
\label{eq:ad-a}A_4=\displaystyle {\sum_{\kappa _{12}\sigma _{12}\kappa
_{12}^{\prime }\sigma _{12}^{\prime }}}\tilde \Theta \left( n_i\lambda
_i,n_j\lambda _j,n_i^{\prime }\lambda _i^{\prime },n_j^{\prime }\lambda
_j^{\prime },\Xi \right) A_8,
\end{equation}
where
\begin{equation}
\label{eq:ad}
\begin{array}[b]{c}
\tilde \Theta \left( n_i\lambda _i,n_j\lambda _j,n_i^{\prime }\lambda
_i^{\prime },n_j^{\prime }\lambda _j^{\prime },\Xi \right) =\frac 12\left(
-1\right) ^{k-p+1+l_i^{\prime }+l_j^{\prime }+\kappa _2+\sigma _2+\kappa
_{12}+\sigma _{12}}\times \\
\times \left[ \kappa _{12},\sigma _{12},\kappa _{12}^{\prime },\sigma
_{12}^{\prime }\right] ^{1/2}\left( n_i\lambda _in_j\lambda _j||g^{\left(
\kappa _1\kappa _2k,\sigma _1\sigma _2k\right) }||n_i^{\prime }\lambda
_i^{\prime }n_j^{\prime }\lambda _j^{\prime }\right) \times \\
\times \left\{
\begin{array}{ccc}
l_i & l_i^{\prime } & \kappa _1 \\
l_j^{\prime } & l_j & \kappa _2 \\
\kappa _{12} & \kappa _{12}^{\prime } & k
\end{array}
\right\} \left\{
\begin{array}{ccc}
s & s & \sigma _1 \\
s & s & \sigma _2 \\
\sigma _{12} & \sigma _{12}^{\prime } & k
\end{array}
\right\} .
\end{array}
\end{equation}

(5) Distributions $\beta \alpha \alpha \alpha $, $\alpha \beta \alpha \alpha $
(cases 7, 8 from Table \ref{israis})
(Figure 1, $A_5$):
\begin{equation}
\label{eq:ae-a}
\begin{array}[b]{c}
A_5=
\displaystyle {\sum_{\kappa _{12}\sigma _{12}\kappa _{12}^{\prime }\sigma
_{12}^{\prime }}}\tilde \Theta \left( n_i\lambda _i,n_j\lambda
_j,n_i^{\prime }\lambda _i^{\prime },n_j^{\prime }\lambda _j^{\prime },\Xi
\right) \times  \\ \times \left[ a^{\left( l_\beta s\right) }\times \left[
a^{\left( l_\alpha s\right) }\times \left[ \tilde a^{\left( l_\alpha
s\right) }\times \tilde a^{\left( l_\alpha s\right) }\right] ^{\left( \kappa
_{12}^{\prime }\sigma _{12}^{\prime }\right) }\right] ^{\left( K_lK_s\right)
}\right] _{p,-p}^{\left( kk\right) },
\end{array}
\end{equation}
where
\begin{equation}
\label{eq;ae-aa}
\begin{array}{c}
\tilde \Theta \left( n_i\lambda _i,n_j\lambda _j,n_i^{\prime }\lambda
_i^{\prime },n_j^{\prime }\lambda _j^{\prime },\Xi \right) \equiv \Theta
^{\prime }\left( n_i\lambda _i,n_j\lambda _j,n_i^{\prime }\lambda _i^{\prime
},n_j^{\prime }\lambda _j^{\prime },\Xi \right) \equiv  \\
\equiv \Theta \left( n_\alpha \lambda _\alpha ,n_\beta \lambda _\beta ,\Xi
\right).
\end{array}
\end{equation}
When $ \widehat{G}(T)=\widehat{G}(\beta \alpha \alpha \alpha )$, then
\begin{equation}
\label{eq:ae-b}
\begin{array}[b]{c}
\tilde \Theta \left( n_i\lambda _i,n_j\lambda _j,n_i^{\prime }\lambda
_i^{\prime },n_j^{\prime }\lambda _j^{\prime },\Xi \right) =\frac 12\left(
-1\right) ^{k-p+\kappa _{12}^{\prime }+\sigma _{12}^{\prime }+l_\alpha
+l_\beta }\left[ \kappa _{12},\sigma _{12}\right] \times  \\
\times \left[ \kappa _{12}^{\prime },\sigma _{12}^{\prime }\right]
^{1/2}\left( n_\beta \lambda _\beta n_\alpha \lambda _\alpha ||g^{\left(
\kappa _1\kappa _2k,\sigma _1\sigma _2k\right) }||n_\alpha \lambda _\alpha
n_\alpha \lambda _\alpha \right) \times  \\
\times \left\{
\begin{array}{ccc}
l_\alpha  & l_\alpha  & \kappa _{12}^{\prime } \\
\kappa _1 & \kappa _2 & k \\
l_\beta  & l_\alpha  & \kappa _{12}
\end{array}
\right\} \left\{
\begin{array}{ccc}
s & s & \sigma _{12}^{\prime } \\
\sigma _1 & \sigma _2 & k \\
s & s & \sigma _{12}
\end{array}
\right\} \times  \\
\times \displaystyle {\sum_{K_lK_s}}\left[ K_l,K_s\right] ^{1/2}\left\{
\begin{array}{ccc}
l_\beta  & l_\alpha  & \kappa _{12} \\
\kappa _{12}^{\prime } & k & K_l
\end{array}
\right\} \left\{
\begin{array}{ccc}
s & s & \sigma _{12} \\
\sigma _{12}^{\prime } & k & K_s
\end{array}
\right\} ,
\end{array}
\end{equation}
and  when $\widehat{G}(T)=\widehat{G}(\alpha \beta \alpha \alpha )$, then
\begin{equation}
\label{eq:ae}
\begin{array}[b]{c}
\tilde \Theta \left( n_i\lambda _i,n_j\lambda _j,n_i^{\prime }\lambda
_i^{\prime },n_j^{\prime }\lambda _j^{\prime },\Xi \right) =\frac 12\left(
-1\right) ^{k-p+\kappa _{12}^{\prime }+\sigma _{12}^{\prime }+\kappa
_{12}+\sigma _{12}}\left[ \kappa _{12},\sigma _{12}\right] \times  \\
\times \left[ \kappa _{12}^{\prime },\sigma _{12}^{\prime }\right]
^{1/2}\left( n_\alpha \lambda _\alpha n_\beta \lambda _\beta ||g^{\left(
\kappa _1\kappa _2k,\sigma _1\sigma _2k\right) }||n_\alpha \lambda _\alpha
n_\alpha \lambda _\alpha \right) \times  \\
\times \left\{
\begin{array}{ccc}
l_\alpha  & l_\alpha  & \kappa _{12}^{\prime } \\
\kappa _1 & \kappa _2 & k \\
l_\alpha  & l_\beta  & \kappa _{12}
\end{array}
\right\} \left\{
\begin{array}{ccc}
s & s & \sigma _{12}^{\prime } \\
\sigma _1 & \sigma _2 & k \\
s & s & \sigma _{12}
\end{array}
\right\} \times  \\
\times \displaystyle {\sum_{K_lK_s}}\left[ K_l,K_s\right] ^{1/2}\left\{
\begin{array}{ccc}
l_\beta  & l_\alpha  & \kappa _{12} \\
\kappa _{12}^{\prime } & k & K_l
\end{array}
\right\} \left\{
\begin{array}{ccc}
s & s & \sigma _{12} \\
\sigma _{12}^{\prime } & k & K_s
\end{array}
\right\} .
\end{array}
\end{equation}

(6) Distributions $\beta \beta \beta \alpha $, $\beta \beta \alpha \beta $
(cases 9, 10 from Table \ref{israis})
(Figure 1, $A_5$):
\begin{equation}
\label{eq:af-a}
\begin{array}[b]{c}
A_5=
\displaystyle {\sum_{\kappa _{12}\sigma _{12}\kappa _{12}^{\prime }\sigma
_{12}^{\prime }}}\tilde \Theta \left( n_i\lambda _i,n_j\lambda
_j,n_i^{\prime }\lambda _i^{\prime },n_j^{\prime }\lambda _j^{\prime },\Xi
\right) \times  \\ \times \left[ \left[ \left[ a^{\left( l_\beta s\right)
}\times a^{\left( l_\beta s\right) }\right] ^{\left( \kappa _{12}\sigma
_{12}\right) }\times \tilde a^{\left( l_\beta s\right) }\right] ^{\left(
K_lK_s\right) }\times \tilde a^{\left( l_\alpha s\right) }\right]
_{p,-p}^{\left( kk\right) },
\end{array}
\end{equation}
where
\begin{equation}
\label{eq:af-aa}
\begin{array}[b]{c}
\tilde \Theta \left( n_i\lambda _i,n_j\lambda _j,n_i^{\prime }\lambda
_i^{\prime },n_j^{\prime }\lambda _j^{\prime },\Xi \right) \equiv \Theta
^{\prime }\left( n_i\lambda _i,n_j\lambda _j,n_i^{\prime }\lambda _i^{\prime
},n_j^{\prime }\lambda _j^{\prime },\Xi \right) \equiv  \\
\equiv \Theta \left( n_\alpha \lambda _\alpha ,n_\beta \lambda _\beta ,\Xi
\right) .
\end{array}
\end{equation}
When $\widehat{G}(T) = \widehat{G}(\beta \beta \beta \alpha )$:
\begin{equation}
\label{eq:af-b}
\begin{array}[b]{c}
\tilde \Theta \left( n_i\lambda _i,n_j\lambda _j,n_i^{\prime }\lambda
_i^{\prime },n_j^{\prime }\lambda _j^{\prime },\Xi \right) =\frac 12\left(
-1\right) ^{k-p+\kappa _{12}^{\prime }+\sigma _{12}^{\prime }+l_\alpha
+l_\beta }\left[ \kappa _{12}^{\prime },\sigma _{12}^{\prime }\right] \times
\\
\times \left[ \kappa _{12},\sigma _{12}\right] ^{1/2}\left( n_\beta \lambda
_\beta n_\beta \lambda _\beta ||g^{\left( \kappa _1\kappa _2k,\sigma
_1\sigma _2k\right) }||n_\beta \lambda _\beta n_\alpha \lambda _\alpha
\right) \times  \\
\times \left\{
\begin{array}{ccc}
l_\beta  & l_\alpha  & \kappa _{12}^{\prime } \\
\kappa _1 & \kappa _2 & k \\
l_\beta  & l_\beta  & \kappa _{12}
\end{array}
\right\} \left\{
\begin{array}{ccc}
s & s & \sigma _{12}^{\prime } \\
\sigma _1 & \sigma _2 & k \\
s & s & \sigma _{12}
\end{array}
\right\} \times  \\
\times \displaystyle {\sum_{K_lK_s}}\left[ K_l,K_s\right] ^{1/2}\left\{
\begin{array}{ccc}
l_\alpha  & l_\beta  & \kappa _{12}^{\prime } \\
\kappa _{12} & k & K_l
\end{array}
\right\} \left\{
\begin{array}{ccc}
s & s & \sigma _{12}^{\prime } \\
\sigma _{12} & k & K_s
\end{array}
\right\} ,
\end{array}
\end{equation}
and when $\widehat{G}(T)=\widehat{G}(\beta \beta \alpha \beta )$:
\begin{equation}
\label{eq:af}
\begin{array}[b]{c}
\tilde \Theta \left( n_i\lambda _i,n_j\lambda _j,n_i^{\prime }\lambda
_i^{\prime },n_j^{\prime }\lambda _j^{\prime },\Xi \right) =\frac 12\left(
-1\right) ^{k-p+\kappa _{12}^{\prime }+\sigma _{12}^{\prime }+\kappa
_{12}+\sigma _{12}}\left[ \kappa _{12}^{\prime },\sigma _{12}^{\prime
}\right] \times  \\
\times \left[ \kappa _{12},\sigma _{12}\right] ^{1/2}\left( n_\beta \lambda
_\beta n_\beta \lambda _\beta ||g^{\left( \kappa _1\kappa _2k,\sigma
_1\sigma _2k\right) }||n_\alpha \lambda _\alpha n_\beta \lambda _\beta
\right) \times  \\
\times \left\{
\begin{array}{ccc}
l_\alpha  & l_\beta  & \kappa _{12}^{\prime } \\
\kappa _1 & \kappa _2 & k \\
l_\beta  & l_\beta  & \kappa _{12}
\end{array}
\right\} \left\{
\begin{array}{ccc}
s & s & \sigma _{12}^{\prime } \\
\sigma _1 & \sigma _2 & k \\
s & s & \sigma _{12}
\end{array}
\right\} \times  \\
\times \displaystyle {\sum_{K_lK_s}}\left[ K_l,K_s\right] ^{1/2}\left\{
\begin{array}{ccc}
l_\alpha  & l_\beta  & \kappa _{12}^{\prime } \\
\kappa _{12} & k & K_l
\end{array}
\right\} \left\{
\begin{array}{ccc}
s & s & \sigma _{12}^{\prime } \\
\sigma _{12} & k & K_s
\end{array}
\right\} .
\end{array}
\end{equation}
The final analytical expressions for diagram $A_6$ appearing in
(\ref{eq:ab-a}), diagram $A_7$ in (\ref{eq:ac-a}) and
diagram $A_8$ in (\ref{eq:ad-a}), are obtained after the following
graphical
transformations:
\begin{enumerate}
\item[ (i)] The second quantization operators are interchanged, until (from left to
right) first come the operators acting upon shell $\alpha $ , then
correspondingly upon $\beta ,$ $\gamma ,$ $\delta $.

\item[(ii)] The generalized Clebsch-Gordan coefficient is transformed to match the
order of operators. This is performed by changing the order of angular
momenta coupling at some of the nodes 1, 2, 3 (Figure 1, $A_6$,
$A_7$ and $A_8$).
\end{enumerate}

We write down immediately the algebraic expressions for
diagrams $A_6$, $A_7$ and $A_8$ (of Figure 1)
after transformation (i) and
(ii) by applying usual generalized graphical technique (see
Gaigalas {\it et al} 1985).
Also we have to notice that $\Theta ^{\prime }$ is equal to $\tilde \Theta $
with a phase factor, which is found by transforming
the diagram of the tensorial
structure according to the rules (i) and (ii). Only in cases 1
and 7-10 (see
Table \ref{israis}) is $\Theta ^{\prime }\equiv \tilde \Theta $,
because there
is no need to transform the tensorial structure.

\begin{center}
*\ \ \ *\ \ \ *
\end{center}

As an example, let us consider in particular the case where operator $a_j$ acts
upon the first shell $n_1\lambda _1$, operator $a_i$ acts upon the second
shell $n_2\lambda _2$, and operators $a_{i^{\prime }}^{\dagger}$,
$a_{j^{\prime
}}^{\dagger}$ act upon the third shell $n_3\lambda _3$ (see Eq. (\ref{eq:pirga})).
This is the distribution 18 in Table \ref{israis}. We obtain the algebraic
expression for distribution $\beta \alpha \gamma \gamma
$ from (\ref{eq:ac-a}). The two-particle operator for this
distribution can be represented by diagram $B_1$ which is
proportional to its tensorial part (diagram $B_2$) as
(Figure 2 $B_1$, $B_2$):

\begin{figure}
\setlength{\unitlength}{1mm}
\begin{picture}(148,110)
\thicklines
%
% Diagram B^{1}
%
\put(20,90){\vector(0,1){10}}
\put(20,100){\vector(0,1){2}}
\put(15,98){\makebox(0,0)[t]{$ n_{2}\lambda_{2}$}}
\put(30,90){\vector(0,1){10}}
\put(30,100){\vector(0,1){2}}
\put(35,98){\makebox(0,0)[t]{$ n_{1}\lambda_{1}$}}
\multiput(21,90)(2,0){5}{\oval(2,1)[t]}
\put(25,93){\makebox(0,0){$kk$}}
\put(20,80){\line(0,1){10}}
\put(20,80){\vector(0,1){3}}
\put(15,83){\makebox(0,0){$ n_{3}\lambda_{3}$}}
\put(20,78){\vector(0,1){3}}
\put(30,80){\line(0,1){10}}
\put(30,80){\vector(0,1){3}}
\put(35,83){\makebox(0,0){$ n_{3}\lambda_{3}$}}
\put(30,78){\vector(0,1){3}}
\put(25,70){\makebox(0,0){$ B_{1}$}}
%
% Diagram B^{2}
%
% linija i
\put(80,90){\vector(0,1){12}}
\put(80,101){\oval(5,5)[b]}
\put(80,90){\circle*{1,7}}
\put(75,98){\makebox(0,0)[t]{$ n_{2}\lambda_{2}$}}
%linja i' (j)
\put(110,90){\line(0,12){12}}
\put(110,102){\vector(0,-1){3}}
\put(110,101){\oval(5,5)[b]}
\put(110,90){\circle*{1,7}}
\put(115,98){\makebox(0,0)[t]{$ n_{3}\lambda_{3}$}}
% forizontali linija desine
\put(80,90.6){\line(6,0){6}}
\put(80,90.3){\line(6,0){6}}
\put(80,90){\line(20,0){30}}
\put(80,89.7){\line(6,0){6}}
\put(80,89.4){\line(6,0){6}}
\put(87,93){\makebox(0,0){$\kappa_{12} \sigma_{12}$}}
% forizontali linija kaire
\put(104,90.6){\line(6,0){6}}
\put(104,90.3){\line(6,0){6}}
\put(104,89.7){\line(6,0){6}}
\put(104,89.4){\line(6,0){6}}
\put(103,93){\makebox(0,0){$\kappa_{12}' \sigma_{12}' $}}
%rezultatinis rangas
\put(95,90){\circle*{1,7}}
\put(95,95){\makebox(0,0){${-}$}}
\put(98,87){\makebox(0,0){${2}$}}
\put(95.6,85){\line(0,1){5}}
\put(95.3,85){\line(0,1){5}}
\put(95,85){\line(0,1){5}}
\put(94.7,85){\line(0,1){5}}
\put(94.4,85){\line(0,1){5}}
\put(95,82){\makebox(0,0){$k k$}}
%linija j (i')
\put(80,80){\line(0,1){10}}
\put(74,84){\makebox(0,0){$ n_{1}\lambda_{1}$}}
\put(75,90){\makebox(0,0){${+}$}}
\put(82,87){\makebox(0,0){${1}$}}
\put(80,80){\vector(0,-1){3}}
\put(80,79){\oval(5,5)[t]}
%linja j'
\put(110,80){\line(0,1){10}}
\put(116,84){\makebox(0,0){$ n_{3}\lambda_{3}$}}
\put(115,90){\makebox(0,0){${-}$}}
\put(108,87){\makebox(0,0){${3}$}}
\put(110,77){\vector(0,1){3}}
\put(110,79){\oval(5,5)[t]}
\put(95,70){\makebox(0,0){$ B_{2}$}}
%
% Diagram B^{3}
%
% linija i
\put(10,40){\vector(0,1){12}}
\put(10,51){\oval(5,5)[b]}
\put(10,40){\circle*{1,7}}
\put(5,48){\makebox(0,0)[t]{$ n_{1}\lambda_{1}$}}
%linja i' (j)
\put(40,40){\line(0,12){12}}
\put(40,52){\vector(0,-1){3}}
\put(40,51){\oval(5,5)[b]}
\put(40,40){\circle*{1,7}}
\put(45,48){\makebox(0,0)[t]{$ n_{3}\lambda_{3}$}}
% forizontali linija desine
\put(10,40.6){\line(6,0){6}}
\put(10,40.3){\line(6,0){6}}
\put(10,40){\line(20,0){30}}
\put(10,39.7){\line(6,0){6}}
\put(10,39.4){\line(6,0){6}}
\put(17,43){\makebox(0,0){$\kappa_{12} \sigma_{12}$}}
% forizontali linija kaire
\put(34,40.6){\line(6,0){6}}
\put(34,40.3){\line(6,0){6}}
\put(34,39.7){\line(6,0){6}}
\put(34,39.4){\line(6,0){6}}
\put(33,43){\makebox(0,0){$\kappa_{12}' \sigma_{12}' $}}
%rezultatinis rangas
\put(25,40){\circle*{1,7}}
\put(25,45){\makebox(0,0){${-}$}}
\put(28,37){\makebox(0,0){${2}$}}
\put(25.6,35){\line(0,1){5}}
\put(25.3,35){\line(0,1){5}}
\put(25,35){\line(0,1){5}}
\put(24.7,35){\line(0,1){5}}
\put(24.4,35){\line(0,1){5}}
\put(25,32){\makebox(0,0){$k k$}}
%linija j (i')
\put(10,30){\line(0,1){10}}
\put(4,34){\makebox(0,0){$ n_{2}\lambda_{2}$}}
\put(5,40){\makebox(0,0){${-}$}}
\put(12,37){\makebox(0,0){${1}$}}
\put(10,30){\vector(0,-1){3}}
\put(10,29){\oval(5,5)[t]}
%linja j'
\put(40,30){\line(0,1){10}}
\put(46,34){\makebox(0,0){$ n_{3}\lambda_{3}$}}
\put(45,40){\makebox(0,0){${-}$}}
\put(38,37){\makebox(0,0){${3}$}}
\put(40,27){\vector(0,1){3}}
\put(40,29){\oval(5,5)[t]}
\put(25,20){\makebox(0,0){$ B_{3}$}}
%
% Diagram B^{4}
%
% linija i
\put(80,40){\vector(0,1){12}}
\put(80,51){\oval(5,5)[b]}
\put(80,40){\circle*{1,7}}
\put(75,48){\makebox(0,0)[t]{$ n_{1}\lambda_{1}$}}
%linja i' (j)
\put(110,40){\line(0,12){12}}
\put(110,52){\vector(0,-1){3}}
\put(110,51){\oval(5,5)[b]}
\put(110,40){\circle*{1,7}}
\put(115,48){\makebox(0,0)[t]{$ n_{3}\lambda_{3}$}}
% forizontali linija desine
\put(80,40.6){\line(6,0){6}}
\put(80,40.3){\line(6,0){6}}
\put(80,40){\line(20,0){30}}
\put(80,39.7){\line(6,0){6}}
\put(80,39.4){\line(6,0){6}}
\put(87,43){\makebox(0,0){$\kappa_{12} \sigma_{12}$}}
% forizontali linija kaire
\put(104,40.6){\line(6,0){6}}
\put(104,40.3){\line(6,0){6}}
\put(104,39.7){\line(6,0){6}}
\put(104,39.4){\line(6,0){6}}
\put(103,43){\makebox(0,0){$\kappa_{12}' \sigma_{12}' $}}
%rezultatinis rangas
\put(95,40){\circle*{1,7}}
\put(95,45){\makebox(0,0){${-}$}}
\put(98,37){\makebox(0,0){${2}$}}
\put(95.6,35){\line(0,1){5}}
\put(95.3,35){\line(0,1){5}}
\put(95,35){\line(0,1){5}}
\put(94.7,35){\line(0,1){5}}
\put(94.4,35){\line(0,1){5}}
\put(95,32){\makebox(0,0){$k k$}}
%linija j (i')
\put(80,30){\line(0,1){10}}
\put(74,34){\makebox(0,0){$ n_{2}\lambda_{2}$}}
\put(75,40){\makebox(0,0){${+}$}}
\put(82,37){\makebox(0,0){${1}$}}
\put(80,30){\vector(0,-1){3}}
\put(80,29){\oval(5,5)[t]}
%linja j'
\put(110,30){\line(0,1){10}}
\put(116,34){\makebox(0,0){$ n_{3}\lambda_{3}$}}
\put(115,40){\makebox(0,0){${-}$}}
\put(108,37){\makebox(0,0){${3}$}}
\put(110,27){\vector(0,1){3}}
\put(110,29){\oval(5,5)[t]}
\put(95,20){\makebox(0,0){$ B_{4}$}}
\end{picture}
\caption{Diagrams for distribution
$a^{(l_2 s)}a^{(l_1 s)} \tilde a^{(l_3 s)} \tilde a^{(l_3 s)}$.
Diagram $B_1$ represents a two-particle operator $\widehat{G}$. Diagrams
$B_2$, $B_3$ and $B_4$ represent graphical transformations.
Diagram $B_2$ represents the tensorial part of the two-particle
operator $B_1$ before transformations, diagram $B_3$ represents
this tensorial part after transformation i), and diagram
$B_4$ represents it after transformation ii).}
\label{bb-b}
\end{figure}
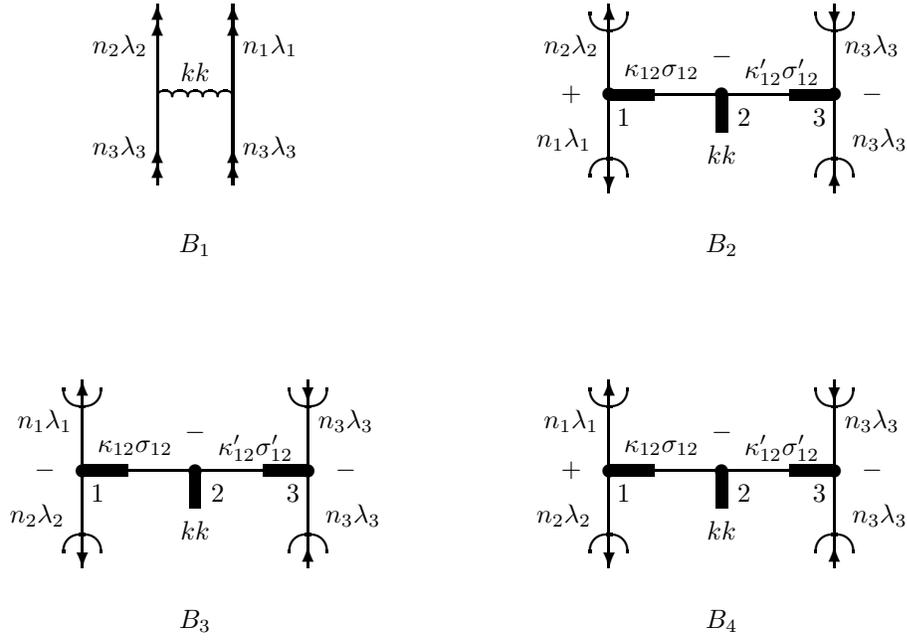

\begin{equation}
\label{eq:ag}B_1=\displaystyle {\sum_{\kappa _{12}\sigma _{12}\kappa
_{12}^{\prime }\sigma _{12}^{\prime }}}\tilde \Theta \left( n_2\lambda
_2,n_1\lambda _1,n_3\lambda _3,n_3\lambda _3,\Xi \right) B_2,
\end{equation}
where
\begin{equation}
\label{eq:ag-aa}
\begin{array}[b]{c}
\tilde \Theta \left( n_2\lambda _2,n_1\lambda _1,n_3\lambda _3,n_3\lambda
_3,\Xi \right) =\frac 12\left( -1\right) ^{k-p+1}\left[ \kappa _{12},\sigma
_{12},\kappa _{12}^{\prime },\sigma _{12}^{\prime }\right] ^{1/2}\times \\
\times \left( n_2\lambda _2n_1\lambda _1||g^{\left( \kappa _1\kappa
_2k,\sigma _1\sigma _2k\right) }||n_3\lambda _3n_3\lambda _3\right) \times
\\
\times \left\{
\begin{array}{ccc}
l_2 & l_3 & \kappa _1 \\
l_1 & l_3 & \kappa _2 \\
\kappa _{12} & \kappa _{12}^{\prime } & k
\end{array}
\right\} \left\{
\begin{array}{ccc}
s & s & \sigma _1 \\
s & s & \sigma _2 \\
\sigma _{12} & \sigma _{12}^{\prime } & k
\end{array}
\right\} .
\end{array}
\end{equation}
We use (i) for diagram $B_2$ (Figure 2, $B_2$) then.
As in the expression (\ref{eq:ag}) the order of second
quantization operators is $a^{\left( l_2s\right) }a^{\left( l_1s\right)
}\tilde a^{\left( l_3s\right) }\tilde a^{\left( l_3s\right) }$, so we
change it according to (i) and obtain
(Figure 2 $B_1$, $B_3$):
\begin{equation}
\label{eq:ah}B_1=-\displaystyle {\sum_{\kappa _{12}\sigma _{12}\kappa
_{12}^{\prime }\sigma _{12}^{\prime }}}\tilde \Theta \left( n_2\lambda
_2,n_1\lambda _1,n_3\lambda _3,n_3\lambda _3,\Xi \right) B_3.
\end{equation}
We use (ii) for diagram $B_3$ then.
We change the sign at the node 1 to obtain finally
(Figure 2 $B_1$, $B_4$):
\begin{equation}
\label{eq:aj}
\begin{array}[b]{c}
B_1=
\displaystyle {\sum_{\kappa _{12}\sigma _{12}\kappa _{12}^{\prime }\sigma
_{12}^{\prime }}}\left( -1\right) ^{l_1+l_2+2s-\kappa _{12}-\sigma
_{12}+1}\times \\ \times \tilde \Theta \left( n_2\lambda _2,n_1\lambda
_1,n_3\lambda _3,n_3\lambda _3,\Xi \right) B_4= \\
=
\displaystyle {\sum_{\kappa _{12}\sigma _{12}\kappa _{12}^{\prime }\sigma
_{12}^{\prime }}}\Theta ^{\prime }\left( n_2\lambda _2,n_1\lambda
_1,n_3\lambda _3,n_3\lambda _3,\Xi \right) B_4= \\ =
\displaystyle {\sum_{\kappa _{12}\sigma _{12}\kappa _{12}^{\prime }\sigma
_{12}^{\prime }}}\Theta ^{\prime }\left( n_2\lambda _2,n_1\lambda
_1,n_3\lambda _3,n_3\lambda _3,\Xi \right) \times \\ \times \left[ \left[
a^{\left( l_1s\right) }\times a^{\left( l_2s\right) }\right] ^{\left( \kappa
_{12}\sigma _{12}\right) }\times \left[ \tilde a^{\left( l_3s\right) }\times
\tilde a^{\left( l_3s\right) }\right] ^{\left( \kappa _{12}^{\prime }\sigma
_{12}^{\prime }\right) }\right] _{p,-p}^{\left( kk\right) },
\end{array}
\end{equation}
where
\begin{equation}
\label{eq:ak}
\begin{array}[b]{c}
\Theta ^{\prime }\left( n_2\lambda _2,n_1\lambda _1,n_3\lambda _3,n_3\lambda
_3,\Xi \right) =\Theta \left( n_1\lambda _1,n_2\lambda _2,n_3\lambda _3,\Xi
\right) = \\
=\left( -1\right) ^{l_1+l_2+2s-\kappa _{12}-\sigma _{12}+1}\tilde \Theta
\left( n_2\lambda _2,n_1\lambda _1,n_3\lambda _3,n_3\lambda _3,\Xi \right) =
\\
=\frac 12
\displaystyle {\sum_{\kappa _{12}\sigma _{12}\kappa _{12}^{\prime }\sigma
_{12}^{\prime }}}\left( -1\right) ^{k-p+l_1+l_2-\kappa _{12}-\sigma
_{12}+1}\left[ \kappa _{12},\sigma _{12},\kappa _{12}^{\prime },\sigma
_{12}^{\prime }\right] ^{1/2}\times \\ \times \left( n_2\lambda _2n_1\lambda
_1||g^{\left( \kappa _1\kappa _2k,\sigma _1\sigma _2k\right) }||n_3\lambda
_3n_3\lambda _3\right) \times \\
\times \left\{
\begin{array}{ccc}
l_2 & l_3 & \kappa _1 \\
l_1 & l_3 & \kappa _2 \\
\kappa _{12} & \kappa _{12}^{\prime } & k
\end{array}
\right\} \left\{
\begin{array}{ccc}
s & s & \sigma _1 \\
s & s & \sigma _2 \\
\sigma _{12} & \sigma _{12}^{\prime } & k
\end{array}
\right\} .
\end{array}
\end{equation}

All the graphical transformations are made and the correspondence between
angular momentum diagrams $B_2$, $B_3$ and $B_4$ in Figure 2 and algebraic expressions is defined according to
the graphical approach of Gaigalas {\it et al} (1985).

Now using (\ref{eq:aj}) and   (\ref{eq:ak}) we
can write down the irreducible tensorial form for
the Coulomb operator with  tensorial structure
$\kappa_1=\kappa_2=$k, $\sigma_1=\sigma_2=0$, $k=0$ and the
two-electron submatrix element:

\begin{equation}
\label{eq:gd-van}
\begin{array}[b]{c}
\left( n_2\lambda _2n_1\lambda _1||g_{Coulomb}^{\left( kk0,000\right)
}||n_3\lambda _3n_3\lambda _3\right)
= \\
=2\left[ k\right] ^{1/2}\left( l_2||C^{\left( k\right) }||l_3
\right) \left( l_1||C^{\left( k\right) }||l_3\right) R_k\left(
n_2l_2n_3l_3,n_1l_1n_3l_3\right) .
\end{array}
\end{equation}

From (\ref{eq:ak}) we have:

\begin{equation}
\label{eq:akivan}
\begin{array}[b]{c}
\Theta ^{\prime }_{Coulomb}
\left( n_2\lambda _2,n_1\lambda _1,n_3\lambda _3,n_3\lambda
_3,\Xi \right) = \\
=\frac 12
\displaystyle {\sum_{\kappa _{12}\sigma _{12}\kappa _{12}^{\prime }\sigma
_{12}^{\prime }}}\left( -1\right) ^{l_1+l_2-\kappa _{12}-\sigma
_{12}+1}\left[ \kappa _{12},\sigma _{12},\kappa _{12}^{\prime },\sigma
_{12}^{\prime }\right] ^{1/2}\times \\
\times
2\left[ k\right] ^{1/2}\left( l_2||C^{\left( k\right) }||l_3
\right) \left( l_1||C^{\left( k\right) }||l_3\right) R_k\left(
n_2l_2n_3l_3,n_1l_1n_3l_3\right)
\times \\
\times \left\{
\begin{array}{ccc}
l_2 & l_3 & k \\
l_1 & l_3 & k \\
\kappa _{12} & \kappa _{12}^{\prime } &0
\end{array}
\right\} \left\{
\begin{array}{ccc}
s & s & 0 \\
s & s & 0 \\
\sigma _{12} & \sigma _{12}^{\prime } &0
\end{array}
\right\} = \\
=-\frac 12 \displaystyle {\sum_{\kappa _{12}\sigma _{12}}}
\left( -1\right) ^{l_2+l_3-\sigma _{12}+k}
\left[ \kappa _{12},\sigma _{12}\right] ^{1/2}\times \\
\times
\left( l_2||C^{\left( k\right) }||l_3
\right) \left( l_1||C^{\left( k\right) }||l_3\right) R_k\left(
n_2l_2n_3l_3,n_1l_1n_3l_3\right)
\left\{
\begin{array}{ccc}
l_2 & l_3 & k \\
l_3 & l_1 & \kappa _{12}
\end{array}
\right\}.
\end{array}
\end{equation}

From (\ref{eq:aj}), by  (\ref{eq:akivan}),
we finally obtain the following expression, in second quantized
form, for the Coulomb operator
for the case $\alpha=1$, $\beta =2$ and $\gamma=3$:

\begin{equation}
\label{eq:ggaa}
\begin{array}[b]{c}
B_{1(Coulomb)}
=-\frac 12
\left( -1\right) ^{l_2+l_3+k}
\left( l_2||C^{\left( k\right) }||l_3
\right) \left( l_1||C^{\left( k\right) }||l_3\right) R_k\left(
n_2l_2n_3l_3,n_1l_1n_3l_3\right)
\times \\
\times
\displaystyle {\sum_{\kappa _{12}\sigma _{12}}}
\left( -1\right) ^{\sigma _{12}}
\left[ \kappa _{12},\sigma _{12}\right] ^{1/2}
\left\{
\begin{array}{ccc}
l_2 & l_3 & k \\
l_3 & l_1 & \kappa _{12}
\end{array}
\right\}
\times \\
\times \left[ \left[ a^{\left( \lambda _1\right) }\times a^{\left( \lambda
_2\right) }\right] ^{\left( \kappa _{12}\sigma _{12}\right) }\times \left[
\stackrel{\sim }{a}^{\left( \lambda _3\right) }\times \stackrel{%
\sim }{a}^{\left( \lambda _3\right) }\right] ^{\left( \kappa
_{12}\sigma _{12}\right) }\right] ^{\left( 00\right) }.
\end{array}
\end{equation}
This kind of operator needs to be calculated when we are
considering, for example, the  matrix element \\
$\left( 3s^23d^2 L_1S_1Q_1L_2 S_2Q_2 LS||H_{Coulomb}||
3s 3d 3p^2 L_1^{\prime}S_1^{\prime}Q_1^{\prime}
L_2^{\prime}S_2^{\prime}Q_2^{\prime}
L_{12}^{\prime}S_{12}^{\prime}
L_3^{\prime}S_3^{\prime}Q_3^{\prime}L^{\prime}S^{\prime} \right)$.

\section*{\bf References}

\begin{quote}
Bar-Shalom A and Klapisch M 1988 {\it Comput. Phys. Commun.} {\bf 50} 375

Burke P G 1970 {\it Comput. Phys. Commun. }{\bf 1} 241

Burke P G, Burke V M and Dunseath K M 1994 {\it J.. Phys. B: At. Mol. Phys.}
{\bf 27} 5341

Condon E U and Shortley G H 1935 {\it The Theory of Atomic Spectra}
(Cambridge: Cambridge University Press)

Cowan R D 1981 {\it The Theory of Atomic Structure and Spectra }(Berkeley,
CA: university of California Press)

Eckart C 1930 {\it Revs. Mod. Phys.} {\bf 2} 305

Fano U 1965 {\it Phys. Rev.} {\bf A67} 140

Gaigalas G A, Kaniauskas J M and Rudzikas Z B 1985 {\it Liet. Fiz. Rink.
(Sov. Phys. Collection)} {\bf 25} 3

Gaigalas G A, Rudzikas Z B and Froese Fischer C 1995 {\it in: Abstracts of
5th EPS (Edited by Dr. R. Pick and G. Thomas)} (Edinburgh UK) 84

Gaigalas G A and Rudzikas Z B 1996 {\it J. Phys. B: At. Mol. Phys.} {\bf 29}
3303

Glass R 1978 {\it Comput. Phys. Commun.} {\bf 16} 11

Glass R and Hibbert A 1978 {\it Comput. Phys. Commun.} {\bf 16} 19

Grant I P 1988 {\it Math. Comput. Chem.} {\bf 2} 1

Jucys A P and Bandzaitis A\ A 1977 {\it Theory of Angular Momentum in
Quantum Mechanics}
(Mokslas: Vilnius) (in Russian).

Jucys A P and Savukynas A J 1973 {\it Mathematical Foundations of the Atomic
Theory} (Vilnius: Mokslas) (in Russian)

Judd B R 1967 {\it Second Quantization and Atomic Spectroscopy} (Baltimore:
John Hopkins Press)

Judd B R 1996 {\it Atomic, Molecular, \& Optical Physics Handbook},
(Edited by G.W.F. Drake) (American Institute of
Physics, Woodbury, New York) p. 88

Lindgren I and Morrison M 1982 {\it Atomic Many-Body Theory 2nd edn
(Springer Series in Chemical Physics 13) }(Berlin: Springer).

Merkelis G\ V, Gaigalas G\ A and Rudzikas Z\ B 1985 {\it Liet. Fiz. Rink.
(Sov. Phys. Coll.) }{\bf 25 }14

Merkelis G\ V and Gaigalas G\ A 1985 {\it Spectroscopy of Autoionized States
of Atoms and Ions }(Moscow: Scientific Council of Spectroscopy) (in Russian)
p. 20

Rudzikas Z B 1991 {\it Comments At. Mol. Phys.} {\bf 26} 269

Rudzikas Z B 1997 {\it Theoretical Atomic Spectroscopy (Many-Electron Atom) }%
(Cambridge: Cambridge University Press) (in press)

Rudzikas Z B and Kaniauskas J M 1984 {\it Quasispin and Isospin in the
Theory of Atom} (Vilnius: Mokslas) (in Russian)

Uylings P\ H\ M 1984 {\it J. Phys. B: At. Mol. Phys.} {\bf 17} 2375

Uylings P\ H\ M 1992 {\it J. Phys. B: At. Mol. Phys.} {\bf 25} 4391

Wigner E P 1931 {\it Gruppentheorie und ihre Anwendung auf die Quantenmechanik
der Atomspektren} (Braunschweig, Friedr. Vieweg)

Yutsis A P, Levinson I B and Vanagas V V 1962 {\it The Theory of Angular
Momentum }(Israel Program for Scientific Translation, Jerusalem)
\end{quote}
\end{document}